\documentclass[10pt,onecolumn,preprintnumbers,amsmath,amssymb]{revtex4}

\usepackage{graphicx}
\usepackage{dcolumn}
\usepackage{bm}
\usepackage{graphicx,epstopdf}
\usepackage{xcolor}
\usepackage{textcomp}
\usepackage{soul}
\usepackage{csquotes}
\usepackage{mathrsfs} 
\usepackage{amsmath}
\usepackage{bigints} 
\usepackage{amsthm,amssymb} 
\usepackage[utf8]{inputenc}
\usepackage[english]{babel}
\usepackage[shortlabels]{enumitem}
\usepackage{mathrsfs,bigints,mathtools,dsfont}
\usepackage[colorlinks=true,linkcolor=blue,citecolor=blue]{hyperref}%

\begin{document}


\title{Extreme events in MLC circuit}

\author{Tapas Kumar Pal}\thanks{Email: pal6590@gmail.com}
\affiliation{Physics and Applied Mathematics Unit, Indian Statistical Institute, Kolkata 700108, India}
\affiliation{Department of Mathematics, Jadavpur University, Kolkata 700032, India}

\author{Dibakar Ghosh}\thanks{Email: dibakar@isical.ac.in}
\affiliation{Physics and Applied Mathematics Unit, Indian Statistical Institute, Kolkata 700108, India}

\begin{abstract}
	
The Murali-Lakshmanan-Chua (MLC) circuit is a well-recognized prominent nonlinear, nonautonomous, and dissipative electronic circuit having a versatile chaotic nature. Unraveling the dynamical synergy responsible for the genesis of \textit{extreme events} in nonlinear dynamical systems is a prolific and spellbinding research area. The present study unveils the dynamical exposition of emerging \textit{extreme events} in the MLC circuit concerning two different \textit{events} being defined in the system. The large expansion of the chaotic attractor following the PM intermittency route plays the crucial role as the precursor behind the emergence of \textit{extreme events} in the system. Our main finding reveals the prevalence of a \textit{force field} due to the presence of externally applied periodic force in the system that creates the dynamical synergy that compels the chaotic trajectory traversing in its phase space to be largely deviated from the residing space, and this large deviation shows the signature of \textit{extreme events}. Apart from the \textit{force field} explication, we explored another two dynamical aspects that also interpret the mechanism behind the genesis of \textit{extreme events} as the large deflection of the chaotic trajectory in the system: the decomposition of the phase space in stable and unstable manifolds concerning slow-fast dynamics and using Floquet multipliers. These two different aspects of calculations of the stable and unstable manifolds explicate the large excursion of the chaotic trajectory as \textit{extreme events} from two different perspectives. We also analyzed the rare occurrences of the \textit{extreme events} statistically using extreme value theory: the threshold \textit{excess values} follow the generalized Pareto distribution, and the \textit{inter-extreme-spike-intervals} follow the generalized extreme value distribution.

\end{abstract}



\maketitle

\section{\label{sec:level}Introduction}

The ubiquitous presence of the inexhaustible comprehensive universe endowed with uncertainty provides birth to versatile, erratic, but vulnerable phenomena. The world, as far as the habitat of the living creatures is concerned, is not beyond the prevalence of these sporadic devastating phenomena. In the pervasiveness of uncertainty, every living creature is adaptively evolved in this world. The predicament situation happens when something abrupt occurs in the living synergy, having a significantly intensified impact and considerably not being a regular or normal phenomenon. The phenomena, not being regular, uncertain to predict, and recurrent but rare to occur spontaneously, having an enormous impact on the living scenario, are generally regarded as \textit{extreme events}\cite{mcphillips2018defining,pal2023extreme,sudharsan2025extreme} (EEs). It is not eloquent to decipher EEs in a pigeonhole; rather, depending on the impact and rarity of occurrence, EEs are defined. Several phenomena extending from natural to anthropogenic have severe aftermaths on human society. The natural catastrophes, such as tsunamis, earthquakes, cyclones, and global warming, to mention a few, and the disasters concerning anthropogenicity, like stock market crashes, blackouts in power grids, nuclear leakage, and many more to mention, have a prodigious impact on the socioeconomy to smooth-functioning the human livelihood. In general a dynamical mechanism\cite{farazmand2019extreme,wu2016universal,chowdhury2022extreme,mehrabbeik2024minimal,rajagopal2024effects} is involved as a precursor behind the genesis of EEs. The survivability and smooth functioning of society claim the exigency of a prolific transdisciplinary research domain\cite{lucarini2016extremes,holland2012extreme,faranda2024statistical,de2024extreme,das2025pattern,das2024complexity,qc99-gh86,chaurasia2020advent,roy2023impact} in this regard concerning prediction\cite{farazmand2016dynamical,meiyazhagan2021prediction,meiyazhagan2021model} as an early warning and the diminution\cite{shashangan2024mitigation,sudharsan2021emergence} to reduce the colossal aftermath of the devastating \textit{extreme} phenomena.

\par In spite of having the metaphysical constraints to define EEs by means of havoc, there are some scientifically well-embraced criteria\cite{ghil2011extreme,chowdhury2022extreme} based on which EEs are categorized in the literature. EEs are inspected and recognized concerning a set of events, either being empirical (observed and collected from real-life scenarios) or generated from a numerical simulation of a dynamical system, as the events lying beyond the range of mathematically calculated standard (existing literal such as standard deviation) measure of dispersion, \textit{i.e.,} \textit{significantly} distant from the central tendency as well as rare in occurrence probabilistically regarding the very specific data set. EEs are being low probable and deviated far from the central tendency; their presence is observed in the tail portion of the skewed non-Gaussian distribution\cite{lucarini2016extremes}. The amount of deflection from the central tendency of the data set signifies the magnitude of impact and the rarity of EEs. As far as the dynamical system is concerned, the chaotic trajectory while traversing in its phase space, if it comes near to or on the unstable (instability region) or repelling region\cite{sapsis2018new} suddenly experiences repulsion that compels the trajectory for ejection in brief from the residing space, and after returns back. This short ejection is reflected in the temporal evolution as high amplitude, potentially indicating EE. The sudden enlargement of the chaotic attractor concerning a dynamical system is the prognostication that EEs may be recognized in the temporal evolution, and so far as the study of \textit{extremes} is concerned, the main two mechanisms responsible for the large expansion of the chaotic attractor are interior crisis\cite{grebogi1987critical,grebogi1982chaotic,grebogi1983crises,ray2019intermittent,ray2020understanding} and intermittency\cite{suresh2020parametric,sudharsan2021symmetrical,wasifextreme}. There are several other mechanisms besides the aforementioned, such as quasiperiodic breakdown\cite{nicolis2006extreme}, transition from librational to rotational motion or vice versa\cite{ghosh2025universal,ray2020extreme,ray2022extreme,pal2023extreme}, attractor hopping in multistable systems due to the presence of noise\cite{kraut2002multistability}, and on-off intermittency\cite{de2013predictability,chowdhury2019synchronization,chowdhury2020distance,nag2021extreme}, to mention a few relevant in the context concerning the generation of EEs.

\par The Muraili-Lakshmanan-Chua (MLC) circuit\cite{thamilmaran2002classification} is a nonautonomous dissipative electronic circuit and is well known because of its versatile chaotic nature following different emerging routes. The rich variety of chaotic nature capacitates the very circuit as a potential candidate for the emergence of extreme events. In this study our main motive is to explore, concerning dynamical sense, whether EEs emerge in this circuit or not, and if they do, what is the crucial dynamical mechanism behind the genesis. We investigated the dynamical nature of the system for two types of \textit{events}, \textit{local minimum} and \textit{local maximum}, of the considered \textit{observable} as the amplitude of the externally applied periodic force is varied. Enthrallingly, we observed the emergence of EEs in both of the two types of \textit{events} following the PM intermittency route of chaos development. Apart from the PM intermittency route of chaos emergence as a precursor to the genesis of EEs, how the EEs are generated in the system is expounded concerning the prevalence of the \textit{force field} due to the presence of the externally applied periodic force, which is one of our main findings. We also explicated the genesis of EEs as the large deflection of the chaotic trajectory in the sense of the \textit{instability} region by decomposing the critical manifold in the \textit{stable} and \textit{unstable} segments concerning the slow-fast dynamics\cite{desroches2012mixed,guckenheimer2008singular,ray2020understanding} being calculated from the autonomous version of the nonautonomous MLC circuit. The discussion of the conversion in the autonomous version of the nonautonomous MLC circuit preserving its dynamical characteristics is illustrated in the main text. Based on the Flouquet multipliers\cite{castelli2013rigorous,castelli2015parameterization}, calculated from the autonomous form of the MLC circuit, the stable and unstable regions of the respective phase space are illustrated, and because of the simultaneous effect of these two manifolds, how the chaotic trajectory becomes largely deviated and shows the signature of EE is interpreted. We also analyzed the EEs, being categorized in the sense of dynamics, statistically\cite{coles2001introduction,lucarini2014towards,sapsis2021statistics} to confirm their rare occurrences regarding two types of non-Gaussian probability distributions based on extreme value theory (EVT): the generalized Pareto distribution (GPD) of the threshold \textit{excess values} and the generalized extreme value (GEV) distribution of the \textit{inter-extreme-spike-intervals}.

\section{\label{sec:leve2}Model Description}

The Murali-Lakshmanan-Chua (MLC) circuit\cite{thamilmaran2002classification} is a familiar and illustrious candidate for being a nonautonomous dissipative simple nonlinear electronic circuit evincing a rich variety of chaotic dynamics. The preliminary focus of this study is a detailed investigation of the MLC circuit concerning whether extreme events (EEs) emerge in the very system in some observable being defined. For the numerical study, the mathematical form governing the MLC circuit variant is presented in the following system:

\begin{equation}
	\begin{aligned}
		\dot{x} &= f \sin(\omega t) - x - y - g(x), \\
		\dot{y} &= \beta x
	\end{aligned}
	\label{eq1}
\end{equation}
where
\[
g(x) = 
\begin{cases}
	bx + (a - b), & x > 1 \\
	ax, & |x| \leq 1 \\
	bx - (a - b), & x < -1
\end{cases}
\]

In our study, specifically, we investigated the dynamical deportment of the system as amplitude $f$ of the externally applied force varies in the range $[0.1312, 0.1315]$. The other parameters are kept fixed as $a=-1.121$, $b=-0.6047$, $\beta=0.05$, and $\omega=0.265$. The numerical simulations are carried out using the fourth-order Runge-Kutta (RK4) algorithm, considering a constant step size of $0.01$, and the initial condition is taken as $(x_{0}, y_{0})=(0.00009, 0.02)$.
\par \textbf{\textit{Categorization of extremes}:} In the literature, the definition of extreme events in concordant is yet to be unsnarled. Commonly, an event is considered as EE if its deviation from the central tendency of the respective data set is significant. The very term \textit{significant} or \textit{significant-deviation} is crucial. There are various techniques available in the literature to substantiate the \textit{significant-deviation}, among which the \textit{significant-height} or \textit{threshold-based} approach is a well-perceived and eulogized one. If $m$ is the mean and $\mu$ is the standard deviation of a data set comprising events, then the events that transcend the threshold $H_{th}=m+d\mu,\,d\in\mathbb{R}\setminus(-1,1)$ are contemplated as EEs. How much an EE is deviated from the central tendency of the corresponding data set elucidating the rareness of the very event is discriminated by the numerical value of $d$; the larger the value of $d$, the more it corroborates the rarity of the event. In this study $x$ is the observable, and two types of events, $x_{max}$ (in the forward time series, the maximum one among the three consecutive $x$ values) and $x_{min}$ (in the forward time series, the minimum one among the three consecutive $x$ values), are considered for meticulous investigation. For the analysis of the events regarding $x_{max}$, the threshold is considered as $H_{th}=m+6\mu$, \textit{i.e.,} corresponding to $d=6$, and for the analysis of events regarding $x_{min}$, the threshold is considered as $H_{th}=m-6\mu$, \textit{i.e.,} corresponding to $d=-6$. In both cases, the events that surpass the threshold either toward the upward direction (for the case of $x_{max}$ events) or toward the downward direction (for the case of $x_{min}$ events) are evinced as EEs.\\

\section{\label{sec:leve3}Genesis of extreme events}

\begin{figure*}[!htbp]
		\centerline{
		\includegraphics[scale = 0.55]{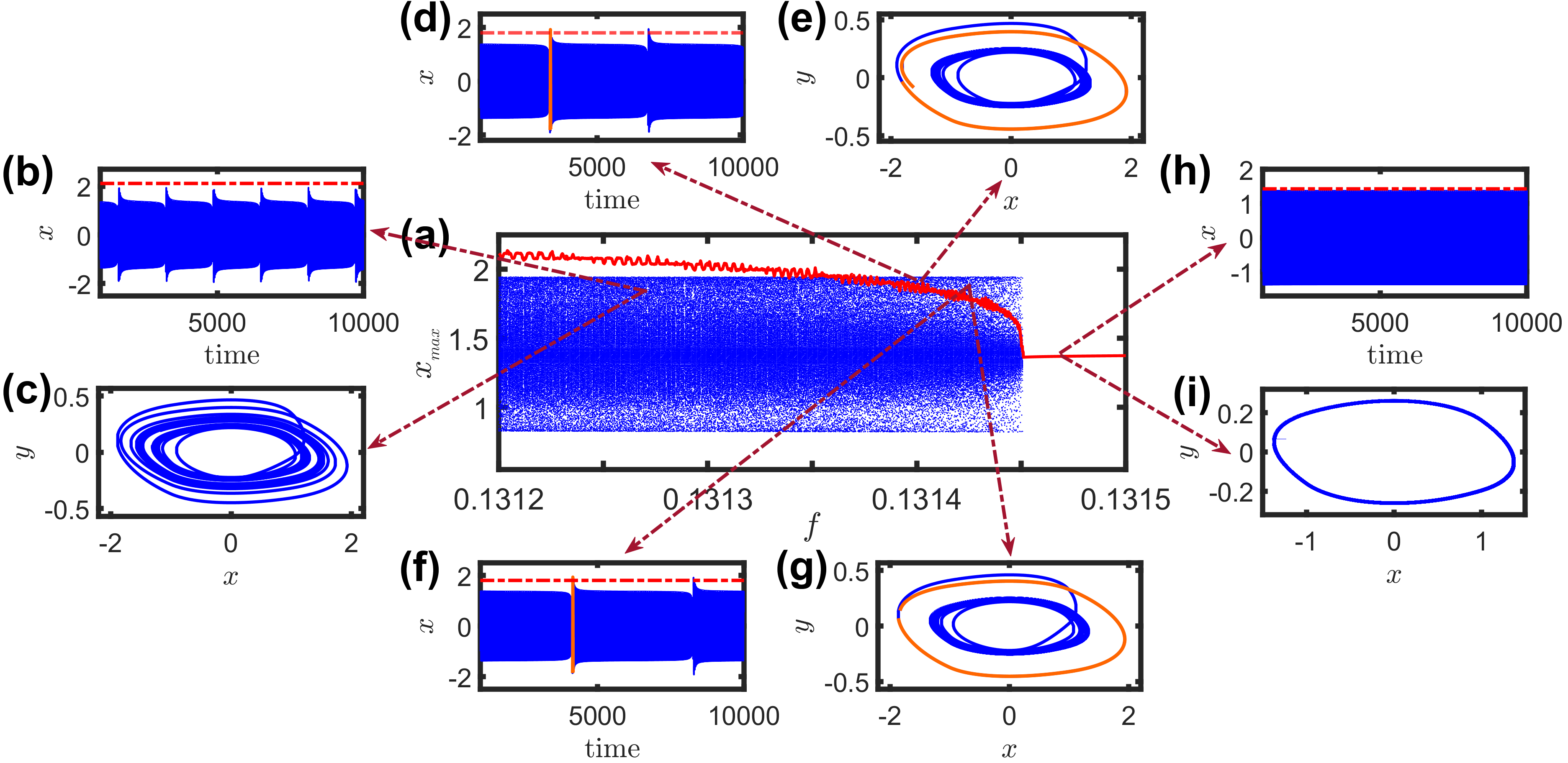}
		
				}
	\caption{\textbf{\textit{Dynamical exposition of the system \eqref{eq1} concerning $x_{max}$}:} (a) The changing portfolio of $x_{max}$ due to the variation of the external forcing amplitude $f$ in the range $[0.1312, 0.1315]$. As $f$ decreases from $0.1315$ following a periodic orbit of one period, large-amplitude chaos emerges through PM intermittency at $f\approx0.131451$. The \textit{red} curve represents the extreme events qualifying threshold $H_{th}=m+6\mu$, where $m$ is the mean and $\mu$ is the standard deviation of the specific data set. The above portion of the threshold curve of the bifurcation diagram showcases the extreme events emerging region. (b) Time series for $f=0.131228$, no spike is observed to surpass the extreme events qualifying \textit{red-dashed} threshold line because the very point is from the non-extreme events chaotic region. (c) Phase space corresponding to $f=0.131228$ upon the $(x,y)$ plane. (d) Time series corresponding to $f=0.1314$. This point is being from the extreme events emerging region; few spikes are observed to cross the \textit{red-dashed} threshold line, which are basically the extreme events. One such extreme event is depicted by the \textit{orange} color, and the respective extreme trajectory is displayed in the phase space diagram upon the $(x,y)$ plane by the \textit{orange} color in (e). (f) Time series corresponding to the point $f=0.131417$. Here also some spikes are observed to exceed the \textit{red-dashed} threshold line, which are extreme events. One such extreme event is depicted by the \textit{orange} color, and the respective extreme trajectory is portrayed by the \textit{orange} color in the phase space diagram upon the $(x,y)$ plane in (g). (h) Time series for $f=0.131463$, which is a periodic time series. The \textit{red-dashed} horizontal line represents the extreme events qualifying threshold. No extreme spike is observed here.(i) Phase space corresponding to $f=0.131463$ on the $(x,y)$ plane. For the detailed numerical calculations, the other parameter values are considered as $a=-1.121$, $b=-0.6047$, $\beta=0.05$, and $\omega=0.265$.
	}  
	\label{fig1}
\end{figure*}

\textbf{\textit{Genesis of extremes concerning the events $x_{max}$}:} The dynamical behavior of the system \eqref{eq1} due to the variation of the amplitude $f$ of the externally applied force, as far as $x_{max}$ is concerned, is meticulously investigated in this section. The changing diegesis of $x_{max}$ as $f$ varies in the range $[0.1312, 0.1315]$ is presented as a bifurcation diagram in Fig. \ref{fig1}(a). A careful observation depicts that as $f$ decreases from $0.1315$ and approaches toward $0.1312$, sudden enlargement of the chaotic attractor from a one-periodic orbit happens via PM intermittency near $f\approx0.131451$. 
This sudden proliferation of the chaotic attractor is the progenitor behind the emergence of extreme events in the very system. The \textit{red} curve demonstrates the extreme events qualifying threshold $H_{th}=m+6\mu$, where $m$ is the mean and $\mu$ is the standard deviation of the corresponding set of events. The above fragment of the \textit{red} threshold curve of the bifurcation diagram is indeed the region where EEs emerge. For a prominent dynamical exposition, four values of $f$ are considered from different regions of the bifurcation diagram: one from the non-extreme chaotic region, $f=0.131228$; two from the extreme events emerging chaotic region, $f=0.1314$ and $f=0.131417$; and one from the non-extreme periodic region, $f=0.131463$, and are studied in detail. Figure \ref{fig1}(b) is the time series of the observable $x$ corresponding to $f=0.131228$, and this value of $f$ being from the non-extreme chaotic region, no spike of the time series is noticed to cross the extreme events qualifying \textit{red-dashed} threshold line $H_{th}$ upward. Figure \ref{fig1}(c) depicts the phase space diagram relating to $f=0.131228$. The temporal evolution of the observable $x$ corresponding to $f=0.1314$ is portrayed in Fig. \ref{fig1}(d). As the value of $f$ is from the extreme events emerging chaotic region of the bifurcation diagram, few spikes of the time series are observed to surpass the extreme events qualifying the \textit{red-dashed} threshold line upward. These spikes are basically the EEs. The temporal evolution of one such extreme spike is indicated by the \textit{orange} color in the temporal dynamics, and the respective extreme trajectory is delineated in the phase space diagram upon the $(x, y)$ plane Fig. \ref{fig1}(e) by the \textit{orange} color. Figure \ref{fig1}(f) discerns the temporal dynamics of the observable $x$ corresponding to $f=0.131417$. This value of $f$ is also from the extreme events emerging chaotic region, so few spikes of the temporal dynamics are observed to exceed the extreme events qualifying \textit{red-dashed} threshold line upward. These spikes are substantially EEs. One such extreme spike is displayed by the \textit{orange} color in the time series, and the respective extreme trajectory is depicted by the \textit{orange} color in the phase space diagram upon the $(x, y)$ plane in Fig. \ref{fig1}(g). The time series of the observable $x$ concerning $f=0.131463$ is shown in Fig. \ref{fig1}(h). Being $f=0.131463$ from the non-extreme periodic region of the bifurcation diagram, 
\begin{figure*}[!htbp]
		\centerline{
		\includegraphics[scale = 0.53]{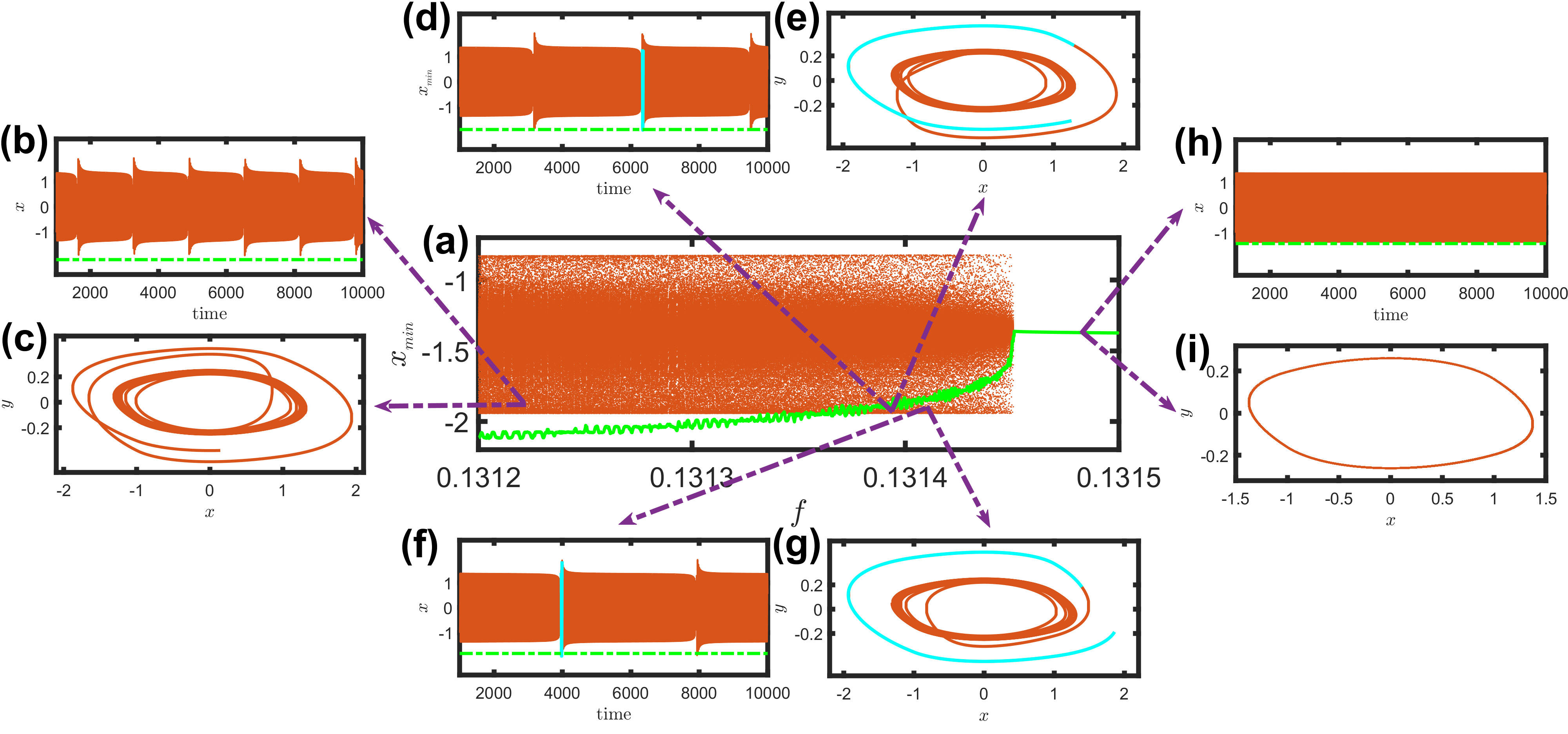}
		
				}
	\caption{\textbf{\textit{Dynamical exposition of the system \eqref{eq1} concerning $x_{min}$}:} (a) The changing framework of $x_{min}$ as $f$ decreases from $0.1315$ to $0.1312$. Large-amplitude chaos is seen to emerge following a one-period periodic orbit at $f\approx0.131451$ through PM intermittency. The \textit{green} curve represents the extreme events qualifying threshold $H_{th}=m-6\mu$, where $m$ is the mean and $\mu$ is the standard deviation of the specific data set. The portion of the bifurcation diagram below the threshold curve $H_{th}$ is the extreme events region. (b) Time series for $f=0.131231$. This point is being from non-extreme chaotic region; no spike is observed to cross the extreme events qualifying \textit{green-dashed} threshold downward. (c) The phase space diagram corresponding to $f=0.131231$. upon $(x,y)$ plane.(d) Time series regarding $f=0.131393$. The very point being from the extreme events qualifying chaotic region, some spikes are observed to exceed the extreme events qualifying \textit{green-dashed} threshold line downward, which are, \textit{au fond,} extreme events. One such extreme event is represented by the \textit{cyan} color spike, and the respective extreme trajectory is displayed by the \textit{cyan} color in the phase space diagram upon the $(x, y)$ plane in (e). (f) Time series corresponding to $f=0.131414$. This point being also from the extreme events qualifying region of the bifurcation diagram, some spikes are observed to cross the extreme event qualifying threshold $H_{th}$ downward, which are extreme events. One extreme event spike is shown by \textit{cyan} color, and the corresponding extreme trajectory is represented by the \textit{cyan} color curve in the phase space diagram on the $(x,y)$ plane in (g). (h) Time series for $f=0.131471$. This point is from the periodic region of the bifurcation diagram. The \textit{green dashed} line represents the extreme events qualifying threshold. No extreme spike is present here. (i) Phase space diagram corresponding to $f=0.131471$ on the $(x,y)$ plane. For the detailed numerical calculations, the other parameter values are considered as $a=-1.121$, $b=-0.6047$, $\beta=0.05$, and $\omega=0.265$.
	}  
	\label{fig2}
\end{figure*}
no spike is seen to exceed the extreme events qualifying \textit{red-dashed} threshold line upward. Figure \ref{fig1}(i) represents the phase space diagram upon the $(x,y)$ plane corresponding to $f=0.131463$. For the bifurcation diagram, the numerical simulation is performed considering $1.5\times10^7$ iterations and abjuring a transient of $1.35\times10^7$, and for all cases of time series and phase spaces, the numerical simulations are performed considering $10^6$ iterations and abjuring a transient of $10^5$.\\

\textbf{\textit{Genesis of extremes concerning the events $x_{min}$}:} The dynamical layout of the system \eqref{eq1} as the amplitude $f$ of the externally applied force varies, regarding $x_{min}$, is scrupulously investigated in this section. The changing framework as $f$ changes in the range $[0.1312, 0.1315]$ is designated in Fig. \ref{fig2}(a). A careful scrutinization unveils, as $f$ decreases from $f=0.1315$ towards $f=0.1312$, a sudden expansion of the chaotic attractor near $f\approx0.131451$ from one-periodic orbit is noticed following PM intermittency. This abrupt expansion of the chaotic attractor is the underlying precursor behind the emergence of EEs in the system \eqref{eq1} concerning $x_{min}$. The \textit{green} curve represents the EEs qualifying threshold $H_{th}=m-6\mu$, where $m$ is the mean and $\mu$ is the standard deviation of the concerning data set. The region of the bifurcation diagram below the \textit{green} curve is \textit{au fond} the EEs emerging region. For a clear understanding of the dynamical portfolio from different regions of the bifurcation diagram, four different values of $f$ are considered: one from the non-extreme chaotic region, $f=0.131231$; two from the extreme events emerging chaotic region, $f=0.131393$ and $f=0.131414$; and one from the non-extreme periodic region, $f=0.131471$, and are studied in detail. The temporal dynamics of the observable $x$ corresponding to $f=0.131231$ is presented in Fig. \ref{fig2}(b). This value of $f$ being from the non-extreme chaotic region of the bifurcation diagram, no spike of the time series is observed to cross the EEs qualifying \textit{green-dashed} threshold line $H_{th}$ downward. Figure \ref{fig2}(c) depicts the phase space upon the $(x, y)$ plane corresponding to $f=0.131231$. The temporal evolution of the observable $x$ regarding $f=0.131393$ is delineated in Fig. \ref{fig2}(d). This value of $f$ being from the EEs qualifying region of the bifurcation diagram, few spikes are noticed to surpass the EEs qualifying \textit{green-dashed} threshold line downward. These spikes are basically the EEs. One such spike is indicated by \textit{cyan} color, and the respective extreme trajectory is shown by \textit{cyan} color in the phase space diagram upon the $(x, y)$ plane in Fig. \ref{fig2}(e). The time series of the observable $x$ corresponding to $f=0.131414$ is displayed in Fig. \ref{fig2}(f). This value of $f$ also being from the EEs qualifying chaotic region of the bifurcation diagram, few spikes are seen to exceed the extreme events qualifying \textit{green-dashed} threshold line downward. These spikes are indeed the EEs. One such spike is indicated by the \textit{cyan} color, and the corresponding extreme trajectory is depicted by the \textit{cyan} color in the phase space diagram upon the $(x, y)$ plane in Fig. \ref{fig2}(g). Figure \ref{fig2}(h) represents the temporal evolution of $x$ concerning $f=0.131471$, which, being from the non-extreme periodic region of the bifurcation diagram, no spike is discerned to cross the extreme events qualifying \textit{green-dashed} threshold line downward. The phase space diagram upon the $(x, y)$ plane corresponding to $f=0.131471$ is presented in Fig. \ref{fig2}(i). For the bifurcation diagram, the numerical simulation is performed considering $1.5\times10^7$ iterations and abjuring a transient of $1.35\times10^7$, and for all cases of time series and phase spaces, the numerical simulations are performed considering $10^6$ iterations and abjuring a transient of $10^5$.\\

\section{\label{sec:leve4} Mechanism behind the genesis of extreme events}

A comprehensive scrutinization concerning the dynamical perspectives lying behind the large excursion of the chaotic trajectory depicting EE is performed in this section. Unveiling the dynamical mechanism responsible for the large excursion of the chaotic trajectory is meticulously discussed here regarding three perspectives: (i) force field, (ii) stable and unstable manifold decomposition using Floquet multipliers, and (iii) slow-fast dynamics. The detailed mathematical calculations of computing the stable (attracting) and unstable (repelling) manifolds concerning the Floquet multipliers as well as regarding the slow-fast dynamics are expounded in the appendix. The chaotic trajectory while wandering in its appurtenant phase space, how it suddenly gets largely deviated, exposing it as \textit{extreme} one from the very phase space, is indeed explicated: using the notion of a force field being prevalent due to the externally applied periodic force; because of the resultant of the simultaneous presence of the stable and unstable manifolds being interpreted using Floquet multipliers\cite{castelli2013rigorous,castelli2015parameterization}; and because of the ejection of the chaotic trajectory from the unstable manifold at the time of entering while strolling over the stable manifold as far as the proximity of the slow-fast dynamics\cite{desroches2012mixed,guckenheimer2008singular,ray2020understanding,farazmand2019extreme} is concerned.

\subsection{Mechanism behind the genesis of extremes concerning the events $x_{max}$}
\label{sec:leve4i}

The detailed elucidation of the origination of EEs concerning the events $x_{max}$ of the system \eqref{eq1} is done in this section. For the detailed explication of the mechanism, the bifurcation parameter value $f=0.1314$ is considered as a specimen in this context.\\

\begin{figure*}[!htbp]
		\centerline{
		\includegraphics[scale = 0.53]{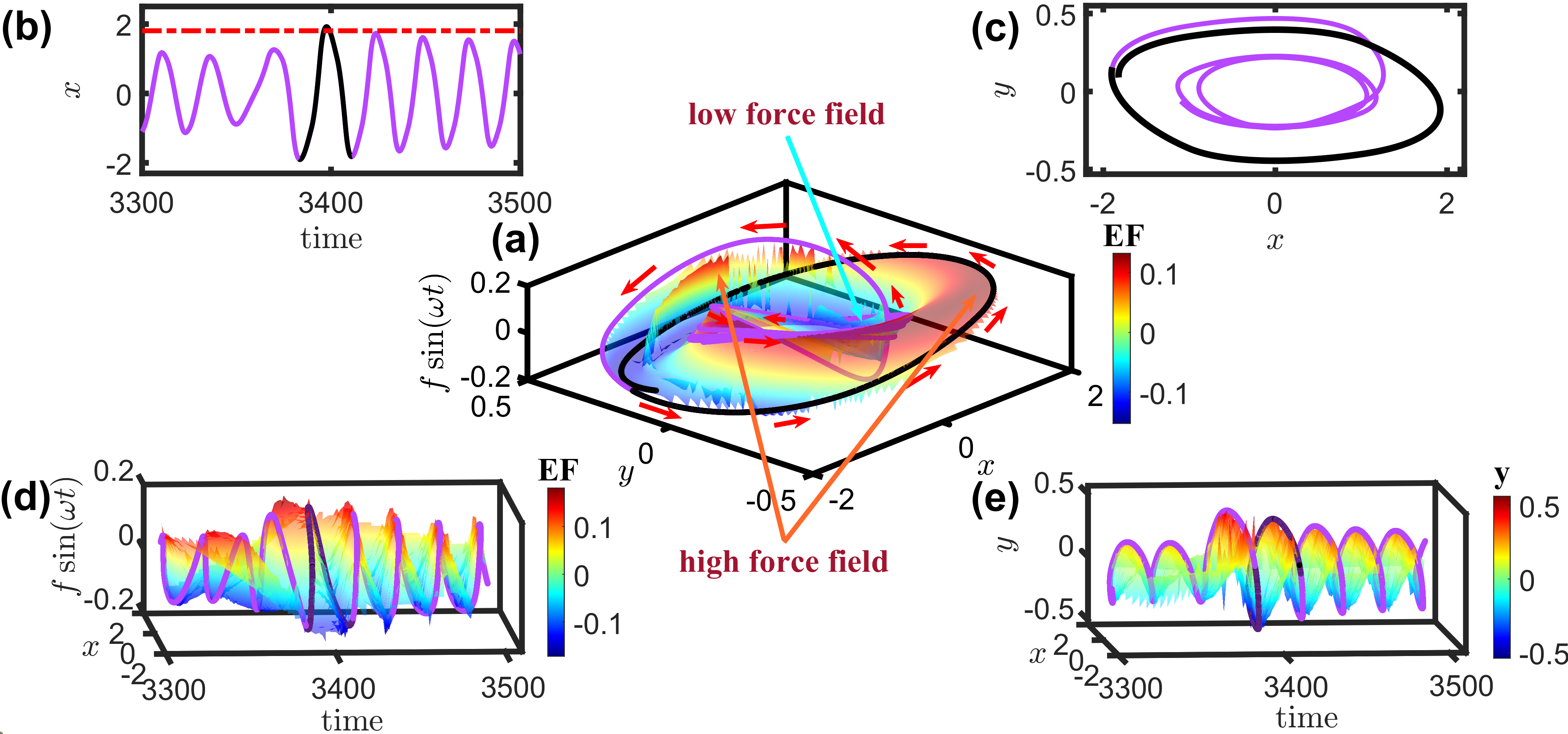}
		
				}
	\caption{\textbf{\textit{Dynamical mechanism behind the large excursion of the chaotic trajectories of the system \eqref{eq1} concerning the events $x_{max}$}:} (a) The synergistic representation of chaotic trajectories and the externally applied periodic force upon the $(x,y.f\sin(\omega t))$ space, elucidating the dynamical mechanism responsible for the significantly large deviation of the chaotic trajectory as EE. The color bar represents the externally applied periodic force. For $f=0.1314$, the chaotic trajectory is depicted by the \textit{violet} color, and it is clearly noticeable from the figure that the trajectory is abruptly pulled out by the high force field from the low force field, and as a result, a large excursion of the chaotic trajectory is recognized, which is basically the EE and is displayed by the \textit{black} color. (b) The time series of the observable $x$ for $f=0.1314$. The \textit{red-dashed} line is the EEs qualifying threshold. One extreme spike of the temporal dynamics is depicted by the \textit{black} color. (c) The phase space upon the $(x,y)$ plane, corresponding to $f=0.1314$. The \textit{black} trajectory represents the EE. (d) The temporal dynamics of the observable $x$ for $f=0.1314$ on the $(t,x,f\sin(\omega t))$ space. The color bar represents the externally applied periodic force. (e) The temporal evolution of the observable $x$ for $f=0.1314$ on the $(t,x,y)$ space. The color bar represents the variable $y$. For the detailed numerical calculations, the other parameter values are considered as $a=-1.121$, $b=-0.6047$, $\beta=0.05$, and $\omega=0.265$.
	}  
	\label{fig15}
\end{figure*}

\textbf{\textit{Force field}:} The presence of the externally applied periodic force $f\sin(\omega t)$ itself develops a force field being shown in Fig. \ref{fig15}(a) on $(x,y,f\sin(\omega t))$ space. The chaotic trajectory corresponding to $f=0.1314$, while strolling on the specified phase space, when it comes near the \textit{significant} low-intensity force field, is suddenly pulled out by the abutting \textit{significant} high-intensity force field, and as a result a large deflection of the chaotic trajectory (shown by the \textit{violet} color) is experienced as EE, which is shown by the \textit{black} color, and the detailed scenario is presented in Fig. \ref{fig15}(a). Figure \ref{fig15}(b) is the depiction of the temporal evolution of the chaotic trajectory; the \textit{red-dashed} line is the EEs qualifying threshold, and the \textit{black} color portion is the respective EE. The projection of the apt phase space on the $(x,y)$ plane is demonstrated in Fig. \ref{fig15}(c); the \textit{black} color portion is the EE. Figure \ref{fig15}(d) represents the elucidation of the temporal dynamics of the observable under the synergistic effect of the externally applied periodic force. The temporal evolution of the observable $x$ upon the $(t,x,y)$ space is delineated in Fig. \ref{fig15}(e). The numerical calculations are performed using the RK4 algorithm, considering $10^6$ iterations and abjuring the transient $10^5$ with a constant step size of $0.01$, and the initial condition is kept fixed as $(x_{0}, y_{0})=(0.00009, 0.02)$.\\

\begin{figure*}[!htbp]
		\centerline{
		\includegraphics[scale = 0.35]{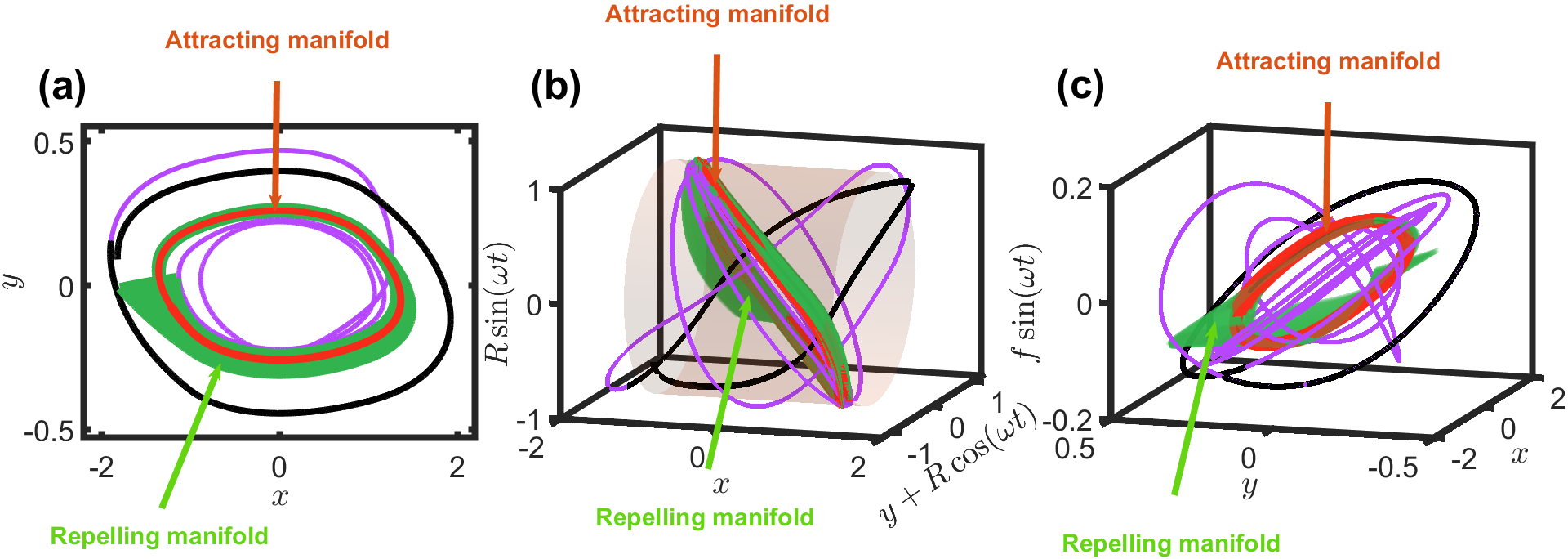}
		
				}
	\caption{\textbf{\textit{The stable and unstable manifold decomposition using Floquet multipliers unravels a responsible mechanism behind the large excursion of the chaotic trajectories of the system \eqref{eq1} concerning the events $x_{max}$}:} For the bifurcation parameter value $f=0.1314$, the genesis of EEs due to the coexistence of the stable (attracting) and the unstable (repelling) manifold is explicated in this panel. (a) The stable manifold is presented by the \textit{red} color, and the unstable manifold is presented by the \textit{green} color on the $(x,y)$ plane. The \textit{violet} color chaotic trajectory represents the phase portrait, and the \textit{black} color represents the \textit{extreme} trajectory. (b) The depiction of the stable and unstable manifold on the \textit{cylindrical} space $(x,y+R\cos(\omega t),R\sin(\omega t))$. The numerical value of $R$ is taken as $1$ here. The \textit{red} color portion represents the stable manifold, and the \textit{green} color portion represents the unstable manifold. The corresponding chaotic trajectory is represented by the \textit{violet} color, and the \textit{black} color portion represents the \textit{extreme} trajectory. (c) The stable and the unstable manifolds are presented by the \textit{red} and \textit{green} colors, respectively, on the $(x,y,f\sin(\omega t))$ space. The \textit{violet} color denotes the corresponding chaotic trajectory, and the \textit{black} color portion represents the EE upon the $(x,y,f\sin(\omega t))$ space. For the detailed numerical calculations, the other parameter values are considered as $a=-1.121$, $b=-0.6047$, $\beta=0.05$, and $\omega=0.265$.
	}  
	\label{fig17}
\end{figure*}

\textbf{\textit{Stable and unstable manifold decomposition using Floquet multipliers}:} The stable and unstable manifolds are calculated based on the Floquet multipliers; the details of the mathematical calculation and the procedure of numerical computation are illustrated in the appendix. The effect of the simultaneous presence of attracting and repelling manifolds on the phase space creates an impact on the chaotic trajectory to largely deviate. Figure \ref{fig17}(a) shows the phase space diagram on the $(x,y)$ plane; the \textit{green} color portion is the repelling manifold, and the \textit{red} color portion is the attracting manifold. The region of presence of the repelling manifold being more than that of the attracting manifold, its effect dominates the phase space, and as a result, the chaotic trajectory, while wandering within the small portion of the phase space, suddenly gets deflected largely and shows the signature of EE as is shown by the \textit{black} color. Figure \ref{fig17}(b) demonstrates the phase space configuration on a \textit{cylindrical} surface. Same as the previous figure, the \textit{green} color segment represents the unstable manifold, and the \textit{red} color portion represents the stable manifold, and as the resultant of the coexistence of these two manifolds, the large deviation of the chaotic trajectory is displayed by the \textit{black} color. Figure \ref{fig17}(c) explicates the phase space layout upon the $(x,y,f\sin(\omega t))$ space. The \textit{green} color section is the unstable manifold, and the \textit{red} color portion is the stable manifold. The large excursion of the chaotic trajectory due to the simultaneous impact of the two manifolds is presented by the \textit{black} color trajectory. The presentations of the phase space and the stable and unstable manifolds in different configurations are only because of clear visualization and recognition from different perspectives. The numerical calculations are performed in MATLAB using the RK4 algorithm, considering $10^6$ iterations and abjuring the transient $10^5$ with a constant step size of $0.01$, and the initial condition is kept fixed as $(x_{0}, y_{0})=(0.00009, 0.02)$.\\ 
  
\begin{figure*}[!htbp]
		\centerline{
		\includegraphics[scale = 0.52]{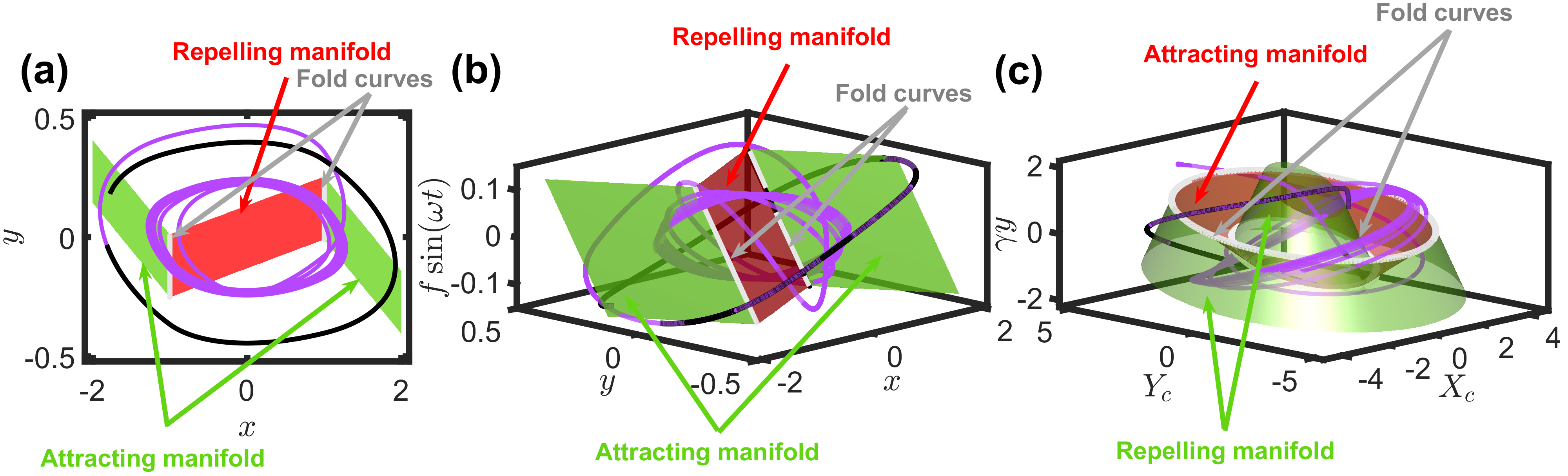}
		
				}
	\caption{\textbf{\textit{The stable and unstable manifold decomposition regarding slow-fast dynamics of the system \eqref{eq1} unravels a responsible mechanism behind the large excursion of the chaotic trajectories concerning the events $x_{max}$}:} Corresponding to the bifurcation parameter value $f=0.1314$, the stable (attracting) and the unstable (repelling) manifolds are calculated regarding the slow-fast dynamics of the system, and due to the simultaneous presence of the stable and the unstable manifolds, how the large excursion of the chaotic trajectory has happened is interpreted in this panel. (a) The stable and the unstable manifolds are presented by the \textit{green} and the \textit{red} colors, respectively, on the $(x,y)$ plane. The respective chaotic trajectory is presented by the \textit{violet} color, and the \textit{black} color portion represents the \textit{extreme} trajectory. The two \textit{gray} lines represents the fold curves. (b) The presentation of the stable and the unstable manifolds on the $(x,y,f\sin(\omega t))$ space. The stable manifold is depicted by the \textit{green} color, and the unstable manifold is shown by the \textit{red} color. The chaotic trajectory is displayed by the \textit{violet} color, and the \textit{black} color portion exhibits the \textit{extreme} trajectory. The two \textit{gray} lines represent the fold curves. (c) The stable manifold is shown by the \textit{red} color, and the unstable manifold is represented by the \textit{green} color (not to be confused with the previous two figures of this panel regarding the colors of the stable and the unstable manifolds) and is presented upon the $(X_{c},Y_{c},\gamma y)$ space. The calculations of $X_{c}$, $Y_{c}$, and $\gamma y$ are discussed in the main text. The presentation configuration is only for the clear visualization in \textit{cylindrical} space. The \textit{gray} color circles represent the fold curves. The respective chaotic trajectory is displayed by the \textit{violet} color, and the \textit{black} color portion of the chaotic trajectory represents the EE. For the detailed numerical calculations, the other parameter values are considered as $a=-1.121$, $b=-0.6047$, $\beta=0.05$, and $\omega=0.265$.
	}  
	\label{fig18}
\end{figure*}

\textbf{\textit{Stable and unstable manifold decomposition concerning slow-fast dynamics}:} The detailed mathematical calculation of how the nonautonomous dynamical system \eqref{eq1} is converted to an autonomous one to discern it in the sense of slow-fast dynamics and to compute the stable and unstable manifolds from the critical manifold of the slow dynamics is engraved in the appendix. Figure \ref{fig18}(a) displays the phase space diagram and the stable and unstable manifolds on the $(x,y)$ plane. The \textit{green} color segment is the stable manifold, and the \textit{red} color section is the unstable manifold. The two \textit{gray} color lines represent the fold curves. The chaotic trajectory, while strolling over the stable (or attracting manifold) on the phase space, when it comes \textit{significantly} close to or on the unstable (or repelling manifold) of the phase space, is immediately ejected, and as a consequence, a large deflection of the chaotic trajectory is noticed as EE, shown by the \textit{black}-colored trajectory in the figure. Figure \ref{fig18}(b) elucidates the stable and unstable manifolds and the phase space configuration on the $(x,y,f\sin(\omega t))$ space. The \textit{green} color segment represents the stable manifold, and the \textit{red} color segment represents the unstable manifold. The two \textit{gray} color lines represent the fold curves.. The chaotic trajectory, while wandering over the stable manifold, when it comes \textit{significantly} near to or on the unstable manifold, gets repulsion for a short while, and as a result, a large excursion of the chaotic trajectory is experienced, which is shown by the \textit{black} color. Figure \ref{fig18}(c) is the illustration of the stable and unstable manifolds on the \textit{cylindrical} phase space. The phase space is drawn upon $(X_{c},Y_{c},\gamma y)$ space, where $X_{c}=(R+\alpha x)\cos\theta$, $Y_{c}=(R+\alpha x)\sin\theta$, and the numerical values of $R$, $\alpha$, and $\gamma$ are taken as $3$, $0.9$, and $5$, respectively. Here $R$ defines the radius of the cylinder, \textit{i.e.,} base radius, and $\alpha$ and $\gamma$ signify, respectively, the radial and vertical scaling. The \textit{green} color surface represents the unstable manifold, and the \textit{red} color surface represents the stable manifold (should not be confused regarding the colors of the stable and unstable manifolds with the other two figures in the same panel). The two \textit{gray} color curves represent the fold curves. The chaotic trajectory, strolling on the stable manifold, when it approaches \textit{significantly} near to or on the unstable manifold, suddenly gets deviated largely, as shown by the \textit{black} color trajectory, which is basically EE. The presentations of the phase space and the stable and unstable manifolds in different configurations are only because of clear visualization and recognition from different perspectives. The numerical calculations are performed in MATLAB using the RK4 algorithm, considering $10^6$ iterations and abjuring the transient $10^5$ with a constant step size of $0.01$, and the initial condition is kept fixed as $(x_{0}, y_{0})=(0.00009, 0.02)$.\\

\subsection{Mechanism behind the genesis of extremes concerning the events $x_{min}$}
\label{sec:leve4ii}

Enthrallingly, in this section while investigating the dynamical mechanism responsible for the large deviation of the chaotic trajectory as EE concerning the events $x_{min}$ of the system \eqref{eq1}, we perceived that it replicates the same mechanism as far as the case of the events $x_{max}$ is concerned. In this context we illustrated the dynamical mechanism behind the large excursion of the chaotic trajectory regarding the events $x_{min}$ for the bifurcation parameter value $f=0.131393$.\\

\begin{figure*}[!htbp]
		\centerline{
		\includegraphics[scale = 0.53]{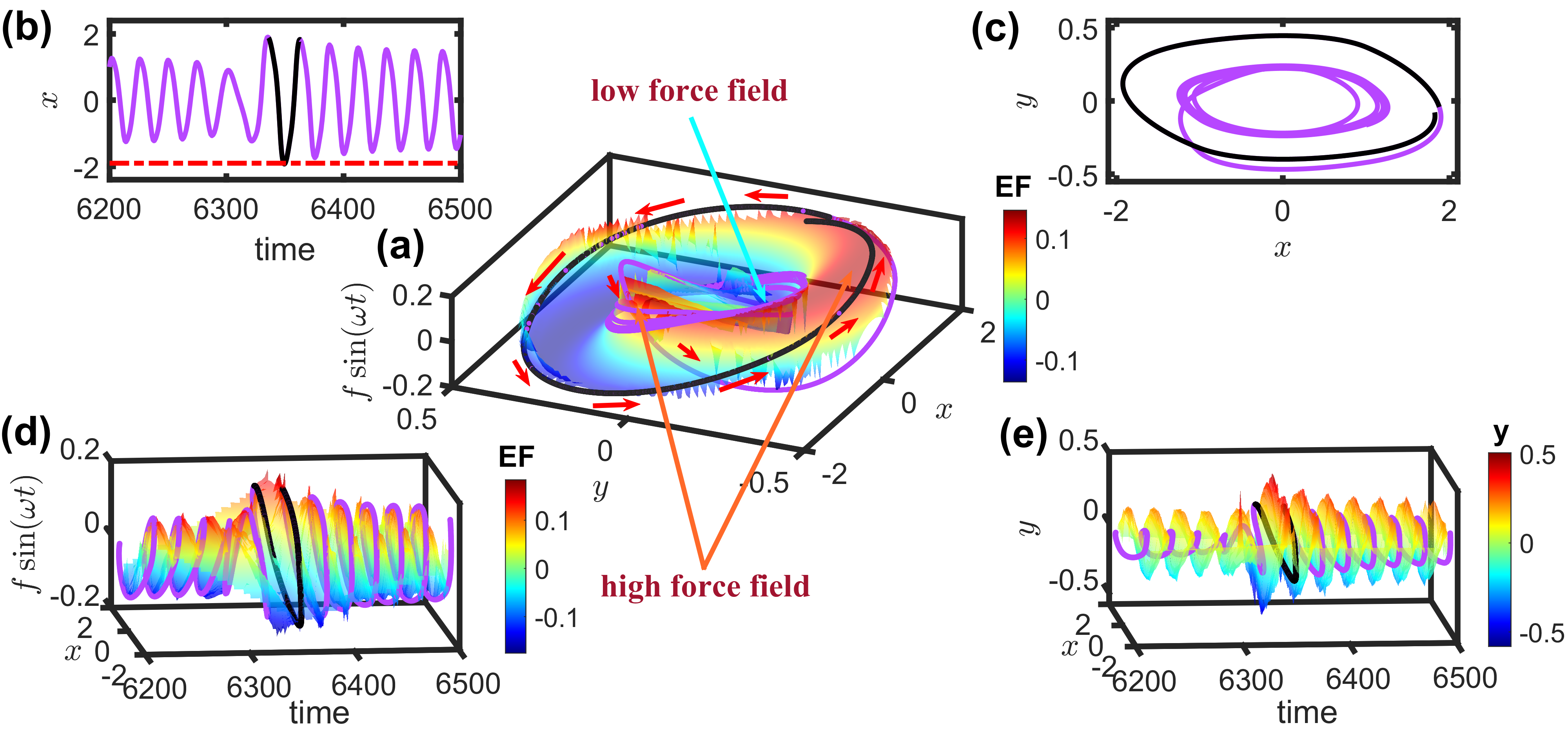}
		
				}
	\caption{\textbf{\textit{Dynamical mechanism behind the large excursion of the chaotic trajectories of the system \eqref{eq1} concerning the events $x_{min}$}:} (a) The dynamical synergy of the chaotic trajectory and the externally applied periodic force is represented upon the $(x,y,f\sin(\omega t))$ plane, explicating the mechanism behind the genesis of the large excursion of the chaotic trajectory depicting EE. The color bar represents the externally applied periodic force. The chaotic trajectory corresponding to $f=0.131393$ is presented by the \textit{violet} color, and it is clearly discernible from the figure that the trajectory is abruptly pulled out by the high force field from the low force field, and as a result, a large deviation of the chaotic trajectory is noticed, which is basically the EE and is displayed by the \textit{black} color. (b) The time series of the observable $x$ for $f=0.131393$. The \textit{red-dashed} line is the EEs qualifying threshold. One downward extreme spike of the temporal dynamics is depicted by the \textit{black} color. (c) The phase space upon the $(x,y)$ plane, corresponding to $f=0.131393$. The \textit{black} trajectory represents the EE. (d) The temporal dynamics of the observable $x$ for $f=0.131393$ on the $(t,x,f\sin(\omega t))$ plane. The color bar represents the externally applied periodic force. (e) The temporal evolution of the observable $x$ for $f=0.131393$ on the $(t,x,y)$ plane. The color bar represents the variable $y$. For the detailed numerical calculations, the other parameter values are considered as $a=-1.121$, $b=-0.6047$, $\beta=0.05$, and $\omega=0.265$.
	}  
	\label{fig16}
\end{figure*}

\textbf{\textit{Force field}:} The chaotic trajectory while wandering on the force field is being prevailed due to the presence of the externally applied periodic force $f\sin(\omega t)$; when it comes near to or at the \textit{significant} low-intensity force field, it is suddenly attracted towards the abutting \textit{significantly} high-intensity force field, and as a result, the large excursion of the chaotic trajectory happens; the very scenario is presented on the $(x,y,f\sin(\omega t))$ space and is depicted in Fig. \ref{fig16}(a). The \textit{violet} color trajectory represents the common chaotic trajectory, and the \textit{black} color segment is EE, and how the large excursion is getting placed is shown by the \textit{oranged}-colored arrows. Figure \ref{fig16}(b) depicts the temporal dynamics; the \textit{red-dashed} line is the EEs qualifying threshold, and the \textit{black} color section is EE. The phase space upon the $(x,y)$ plane is demonstrated in Fig. \ref{fig16}(c). The \textit{black} color segment is EE. Figure \ref{fig16}(d) elucidates the temporal evolution of the observable considering the synergistic effect of the externally applied periodic force and is presented upon the $(t,x,f\sin(\omega t))$ space. The \textit{black} color segment is EE. Figure \ref{fig16}(e) represents the temporal dynamics of the observable $x$ on the $(t,x,y)$ space. The \textit{black} color section depicts EE. The numerical calculations are performed using the RK4 algorithm, considering $10^6$ iterations and abjuring the transient $10^5$ with a constant step size of $0.01$, and the initial condition is kept fixed as $(x_{0}, y_{0})=(0.00009, 0.02)$.\\

\begin{figure*}[!htbp]
		\centerline{
		\includegraphics[scale = 0.35]{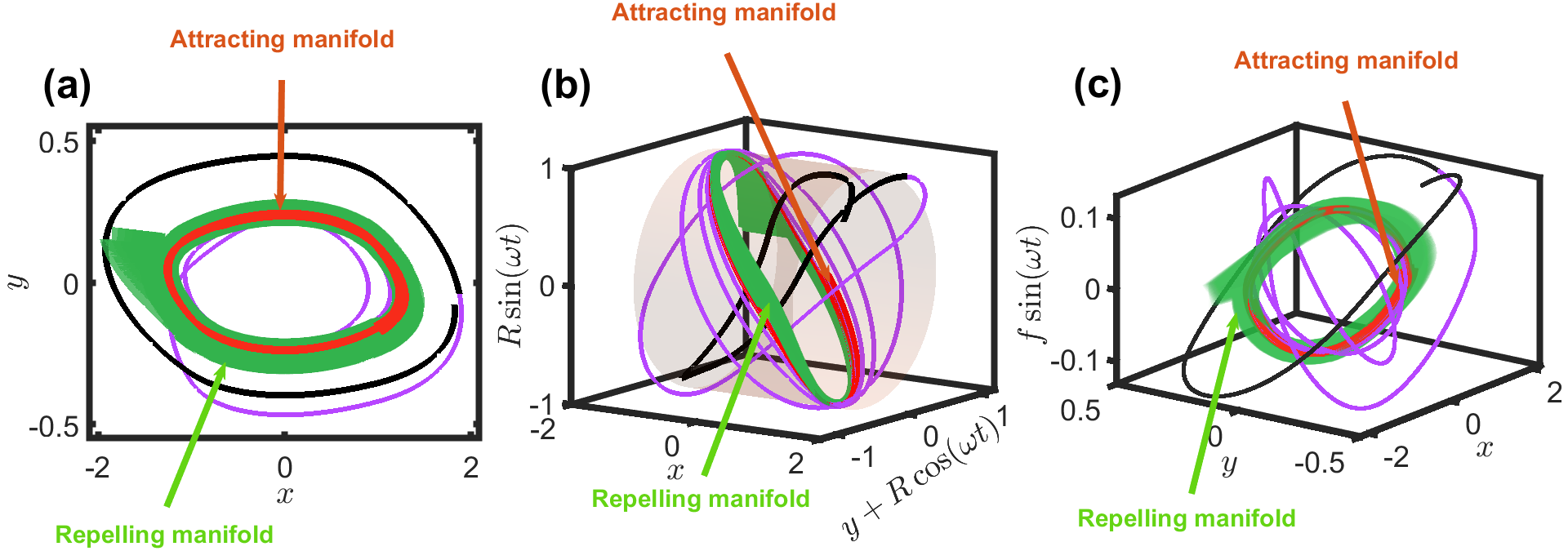}
		
				}
	\caption{\textbf{\textit{The stable and unstable manifold decomposition using Floquet multipliers unravels a responsible mechanism behind the large excursion of the chaotic trajectories of the system \eqref{eq1} concerning the events $x_{min}$}:} For the bifurcation parameter value $f=0.131393$, the genesis of EEs due to the coexistence of the stable (attracting) and the unstable (repelling) manifold is explicated in this panel. (a) The stable manifold is presented by the \textit{red} color, and the unstable manifold is presented by the \textit{green} color on the $(x,y)$ plane. The \textit{violet} color chaotic trajectory represents the phase portrait, and the \textit{black} color represents the \textit{extreme} trajectory. (b) The depiction of the stable and unstable manifold on the \textit{cylindrical} space $(x,y+R\cos(\omega t),R\sin(\omega t))$. The numerical value of $R$ is taken as $1$ here. The \textit{red} color portion represents the stable manifold, and the \textit{green} color portion represents the unstable manifold. The corresponding chaotic trajectory is represented by the \textit{violet} color, and the \textit{black} color portion represents the \textit{extreme} trajectory. (c) The stable and the unstable manifolds are presented by the \textit{red} and \textit{green} colors, respectively, on the $(x,y,f\sin(\omega t))$ space. The \textit{violet} color denotes the corresponding chaotic trajectory, and the \textit{black} color portion represents the EE upon the $(x,y,f\sin(\omega t))$ space. For the detailed numerical calculations, the other parameter values are considered as $a=-1.121$, $b=-0.6047$, $\beta=0.05$, and $\omega=0.265$.
	}  
	\label{fig19}
\end{figure*}

\textbf{\textit{Stable and unstable manifold decomposition using Floquet multipliers}:} Transforming the nonautonomous system \eqref{eq1} to an autonomous one, the stable and unstable manifolds are calculated. The details of the mathematical calculations and the procedure of the numerical computations are illustrated in the appendix. The resultant of the coexistence of the stable and unstable manifolds on the phase space generates an impact, because of which the large deviation of the chaotic trajectory happens. Figure \ref{fig19}(a) is the representation of the phase space along with the stable and unstable manifolds upon the $(x,y)$ plane. The \textit{green} color section demonstrates the repelling (or unstable) manifold, and the \textit{red} color section demonstrates the attracting (or stable) manifold. As the region of presence of the unstable manifold is greater than the region of presence of the stable manifold, the unstable segment dominates the dynamics, and as a consequence, the chaotic trajectory, while wandering on the small portion of the phase space, abruptly gets repulsion and depicts large excursions. The large deviation of the chaotic trajectory is shown by the \textit{black} color. Figure \ref{fig19}(b) explicates the phase space of the chaotic trajectory and the coexisting stable and unstable manifolds on a \textit{cylinder} upon $(x,y+R\cos\theta, R\sin\theta)$ space. The \textit{green} color segment is the unstable manifold, and the \textit{red} color section is the stable manifold, and because of the simultaneous impact of these two manifolds, the large deviation of the chaotic trajectory is depicted by the \textit{black} color segment. Figure \ref{fig19}(c) represents the phase space diagram of the chaotic trajectory along with the stable and unstable manifold upon the $(x,y,f\sin(\omega t))$ space. The \textit{green} color segment is the unstable manifold, and the \textit{red} color segment is the stable manifold. The largely deviated EE trajectory is demonstrated by the \textit{black} color. The presentations of the phase space and the stable and unstable manifolds in different configurations are only because of clear visualization and recognition from different perspectives. The numerical calculations are performed in MATLAB using the RK4 algorithm, considering $10^6$ iterations and abjuring the transient $10^5$ with a constant step size of $0.01$, and the initial condition is kept fixed as $(x_{0}, y_{0})=(0.00009, 0.02)$.\\

\begin{figure*}[!htbp]
		\centerline{
		\includegraphics[scale = 0.52]{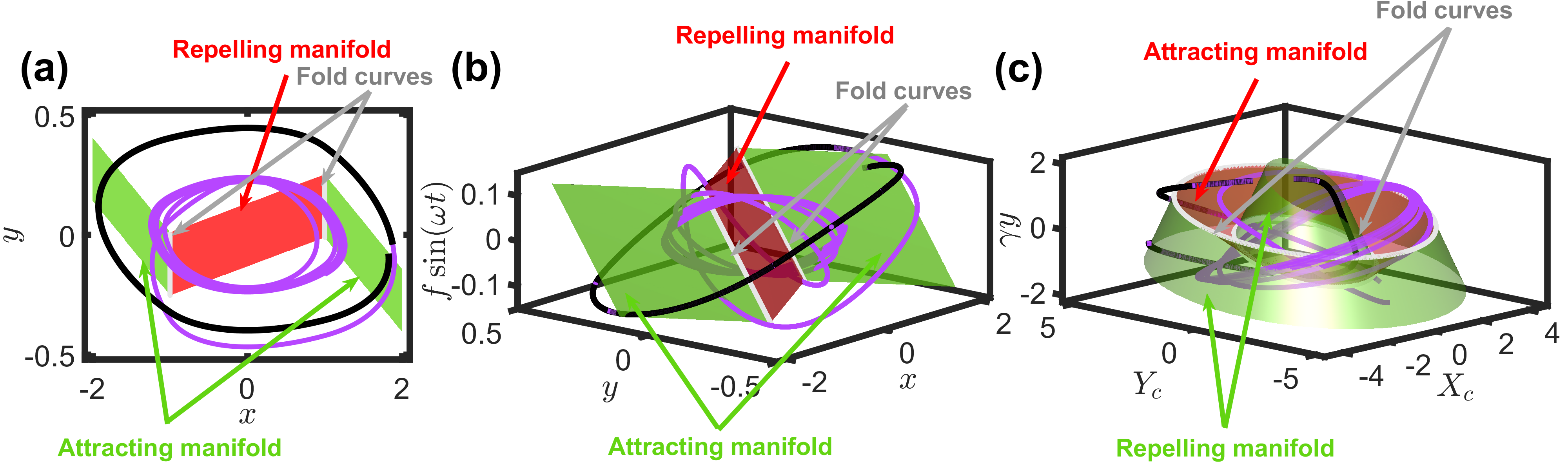}
		
				}
	\caption{\textbf{\textit{The stable and unstable manifold decomposition regarding slow-fast dynamics of the system \eqref{eq1} unravels a responsible mechanism behind the large excursion of the chaotic trajectories concerning the events $x_{min}$}:} Corresponding to the bifurcation parameter value $f=0.131393$, the stable (attracting) and the unstable (repelling) manifolds are calculated regarding the slow-fast dynamics of the system, and due to the simultaneous presence of the stable and the unstable manifolds, how the large excursion of the chaotic trajectory has happened is interpreted in this panel. (a) The stable and the unstable manifolds are presented by the \textit{green} and the \textit{red} colors, respectively, on the $(x,y)$ plane. The respective chaotic trajectory is presented by the \textit{violet} color, and the \textit{black} color portion represents the \textit{extreme} trajectory. The two \textit{gray} lines represents the fold curves. (b) The presentation of the stable and the unstable manifolds on the $(x,y,f\sin(\omega t))$ space. The stable manifold is depicted by the \textit{green} color, and the unstable manifold is shown by the \textit{red} color. The chaotic trajectory is displayed by the \textit{violet} color, and the \textit{black} color portion exhibits the \textit{extreme} trajectory. The two \textit{gray} lines represent the fold curves. (c) The stable manifold is shown by the \textit{red} color, and the unstable manifold is represented by the \textit{green} color (not to be confused with the previous two figures of this panel regarding the colors of the stable and the unstable manifolds) and is presented upon the $(X_{c},Y_{c},\gamma y)$ space. The calculations of $X_{c}$, $Y_{c}$, and $\gamma y$ are discussed in the main text. The presentation configuration is only for the clear visualization in \textit{cylindrical} space. The \textit{gray} color circles represent the fold curves. The respective chaotic trajectory is displayed by the \textit{violet} color, and the \textit{black} color portion of the chaotic trajectory represents the EE. For the detailed numerical calculations, the other parameter values are considered as $a=-1.121$, $b=-0.6047$, $\beta=0.05$, and $\omega=0.265$.
	}  
	\label{fig20}
\end{figure*}

\textbf{\textit{Stable and unstable manifold decomposition concerning slow-fast dynamics}:} The detailed illustration of the mathematical calculations and the numerical computations of the stable and unstable segments of the critical manifold of the slow dynamics of the transformed autonomous system of the nonautonomous system \eqref{eq1} in the sense of slow-fast dynamics is done in the appendix. Figure \ref{fig20}(a) is the representation of the phase space together with the attracting (or stable) and repelling (or unstable) manifolds upon the $(x,y)$ plane. The \textit{green} color portion depicts the stable manifold, and the \textit{red} color portion depicts the unstable manifold. The two \textit{gray} color lines represent the fold curves. The chaotic trajectory, while wandering upon the stable manifold when it approaches \textit{significantly} near to or on the unstable manifold, experiences an abrupt ejection, and as a result, the large excursion happens. The largely deviated \textit{extreme} trajectory is shown by the \textit{black} color. Figure \ref{fig20}(b) elucidates the same mechanism upon the $(x,y,f\sin(\omega t))$ space. The \textit{green} color segment is the stable manifold, and the \textit{red} color segment is the unstable manifold. The two \textit{gray} color lines represent the fold curves. The chaotic trajectory, while wandering on the stable manifold, when it comes close to or on the unstable manifold, immediately gets repulsion, and as a consequence, the large excursion of the trajectory happens. The largely deviated \textit{extreme} trajectory is shown by the \textit{black} color. Figure \ref{fig20}(c) represents the phase space diagram of the chaotic trajectory along with the stable and unstable manifolds in \textit{cylindrical} configuration on the $(X_{c},Y_{c},\gamma y)$ space, where $X_{c}=(R+\alpha x)\cos\theta$, $Y_{c}=(R+\alpha x)\sin\theta$, and the numerical values of $R$, $\alpha$, and $\gamma$ are considered as $3$, $0.9$, and $5$, respectively. Here $R$ defines the radius of the cylinder, \textit{i.e.,} base radius, and $\alpha$ and $\gamma$ signify, respectively, the radial and vertical scaling. The stable and unstable manifolds are demonstrated by the \textit{red} and \textit{green} colors, respectively (should not be confused regarding the colors of the stable and the unstable manifolds with the other two figures in the same panel). Most of the time, the chaotic trajectory resides on the stable manifold, and while strolling, when it comes \textit{significantly} close to or on the unstable manifold, it gets repulsion, and as a result, a large deflection of the trajectory happens. The largely deviated \textit{extreme} chaotic trajectory is depicted by the \textit{black} color. The presentations of the phase space and the stable and unstable manifolds in different configurations are only because of clear visualization and recognition from different perspectives. The two \textit{gray} color curves represent the fold curves. The numerical calculations are performed in MATLAB using the RK4 algorithm, considering $10^6$ iterations and abjuring the transient $10^5$ with a constant step size of $0.01$, and the initial condition is kept fixed as $(x_{0}, y_{0})=(0.00009, 0.02)$.\\

\section{\label{sec:leve5} Statistical analysis of extreme events}

\begin{figure}[!htbp]
	\centering
	\includegraphics[scale=0.265]{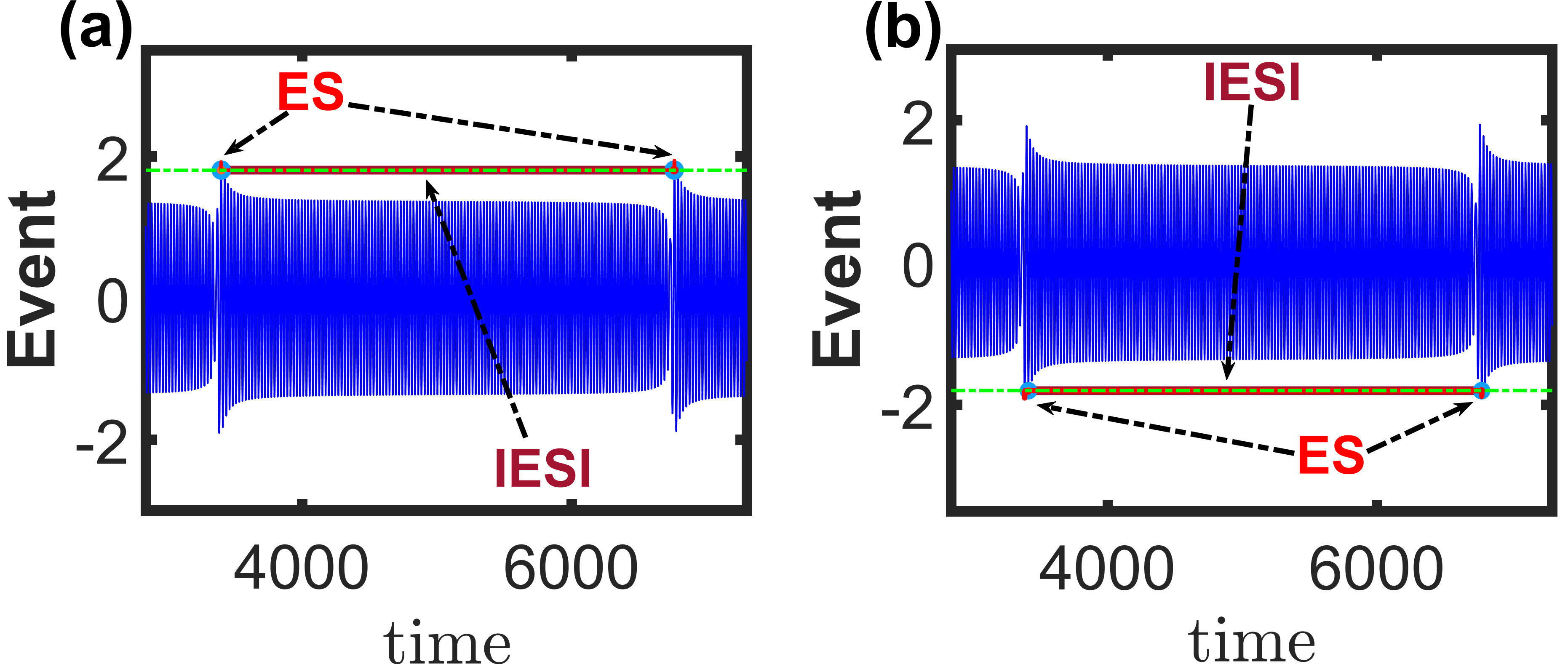} 
	\caption{\textbf{\textit{Schematic presentation of extreme spike (ES) and inter-extreme-spike-interval (IESI)}:} (a)-(b) The \textit{green-dashed} line represents the extreme events qualifying threshold line. The spikes of the time series that are being exceeded by the threshold line are termed as extreme spikes (ES), shown by the arrow in the figures. The elapsed time difference between two consecutive ESs is termed as the \textit{inter-extreme-spike-interval} (IESI), elucidated by the \textit{brown} color line in the figures.
	} 
	\label{fig7}
\end{figure}

A brief statistical description of our detailed study is presented in this section, mainly focusing on extreme value theory (EVT)\cite{lucarini2012universal,felici2007extreme,gupta2011extreme}. At the very outset of each section, a detailed analysis of the histograms concerning the respective events elucidating its explication is articulated. The probability distributions of the threshold \textit{excess values} and the \textit{inter-extreme-spike-intervals} (IESIs) regarding different events are scrutinized meticulously. Threshold \textit{excess value}: The absolute numerical value of the difference of the numerical value of the EE and the numerical value of the EE qualifying threshold is defined as threshold \textit{excess value}. \textit{Inter-extreme-spike-interval} (IESI): The time difference elapsed between two consecutive extreme spikes (ESs), the spikes in the respective time series that exceeded the EE qualifying threshold line, in the forward time series as presented in the schematic diagram in Fig. \ref{fig7} is contemplated as IESI. 

In each case of the analysis regarding each event, the threshold \textit{excess values} follow the generalized Pareto distribution (GPD) described by the following mathematical presentation:

\begin{equation}
	\begin{aligned}
		H(x) &= 1 - \left( 1 + \frac{\alpha x}{\hat{\gamma}} \right)^{-1/\alpha},
		\quad \text{for } \alpha \neq 0, \\[6pt]
		H(x) &= 1 - \exp\!\left( -\frac{x}{\hat{\gamma}} \right),
		\quad \text{for } \alpha = 0.
	\end{aligned}
	\label{eq3}
\end{equation}
\\
where $1 + \dfrac{\alpha x}{\hat{\gamma}} > 0$, $-\infty < \alpha < \infty$, $x > 0$, and $\hat{\gamma} > 0$. The parameters $\alpha$ and $\hat{\gamma}$ are called the shape parameter and the scale parameter, respectively. The tail of the GPD distribution decays exponentially if $\alpha=0$, decays polynomially if $\alpha>0$, and has an upper bound if $\alpha<0$. 

The IESIs in both the cases concerning each event follow the generalized extreme value (GEV) distribution\cite{faranda2011numerical}, being articulated in the following mathematical exposition:

\begin{equation}
	G(x) = \frac{1}{\beta}
	\exp\!\left[
	- \left( 1 + \gamma \frac{x - \alpha}{\beta} \right)^{-\frac{1}{\gamma}}
	\right]
	\left( 1 + \gamma \frac{x - \alpha}{\beta} \right)^{-\frac{1}{\gamma}-1}
	\label{eq4}
\end{equation}
\\
Where $1 + \gamma \frac{x - \alpha}{\beta} > 0$, $\alpha\, (\neq 0)$ is the location parameter, $\beta\, (> 0)$ is the scale parameter, and $\gamma\, (\neq 0)$ is the shape parameter. Depending on the sign of the shape parameter $\gamma$, the distribution is classified in two categories: (i) Fr\'echet (type II) if $\gamma>0$, being bounded below by $0$, and (ii) Weibull (type III) if $\gamma<0$, having an upper bound. There is another case scenario corresponding to $\gamma = 0$, the Gumbel (type I) distribution being unbounded, and its mathematical form is presented in the following:

\begin{equation}
	G(x) = \frac{1}{\beta}
	\exp\!\left(
	- \exp\!\left( - \frac{x - \alpha}{\beta} \right)
	- \frac{x - \alpha}{\beta}
	\right)
	\label{eq5}
\end{equation}
\\

\subsection{Statistical analysis concerning the events $x_{\max}$}
\label{sec:leve5i}

This section is mainly devoted to the description of critical statistical analysis of the system \eqref{eq1} regarding the events $x_{max}$, focusing on various aspects. At the beginning of the section, a brief overview of the histogram presentation of the events, elucidating its explication, is described. Secondly, statistical analysis of the threshold \textit{excess values} 
concerning the probability distribution is designated, and lastly, the statistical analysis of the \textit{inter-extreme-spike-intervals} (IESIs) is described.\\

\textbf{\textit{Histogram presentation}:} The histogram diagrams of the events $x_{max}$ for different values of $f$ are described in this section. In each of the histogram 
\begin{figure*}[!htbp]
		\centerline{
		\includegraphics[scale = 0.56]{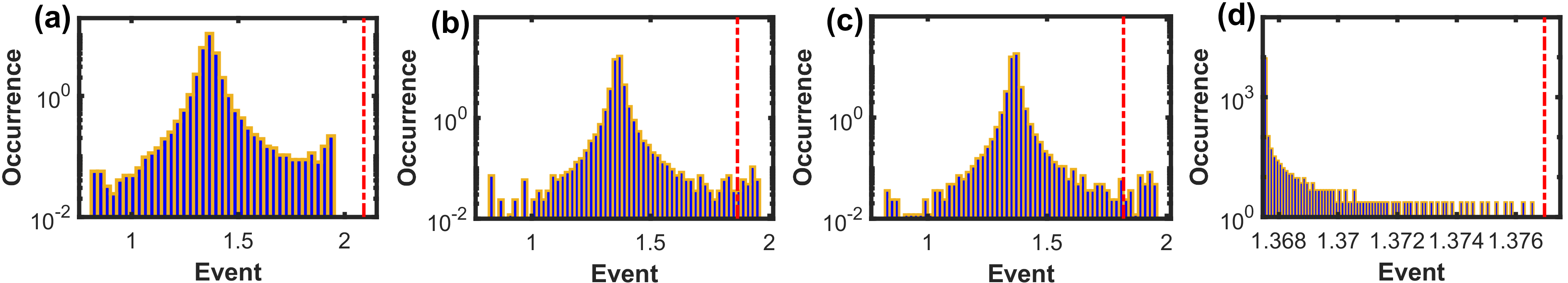}
		
				}
	\caption{\textbf{\textit{Histogram of the events $x_{_{max}}$ for different values of the bifurcation parameter $f$ of the system \eqref{eq1} in semi-log scale}:} The vertical \textit{red-dashed} line represents the extreme events qualifying threshold line $H_{th}=m+6\mu$, where $m$ is the mean and $\mu$ is the standard deviation of the respective sets of data, in each figure for the respective value of $f$. (a) Histogram of the events $x_{_{max}}$ corresponding to $f=0.131228$. No portion of the histogram is observed to surpass the threshold line towards the right side; \textit{i.e.,} no extreme event is present here, because the point is from the non-extreme events chaotic region of the bifurcation diagram. (b) Histogram of the events for $f=0.1314$, which is a point from the extreme events qualifying chaotic region of the bifurcation diagram. A portion of the histogram is observed to cross the extreme events qualifying \textit{red-dashed} vertical threshold line towards the right. This portion au fond elucidates the extreme events. (c) Histogram of the events corresponding to $f=0.131417$. The very point being from the extreme events qualifying region of the bifurcation diagram, a portion of the histogram is seen to exceed the extreme events qualifying \textit{red-dashed} vertical threshold line towards the right. This portion basically depicts the extreme events. (d) Histogram of the events regarding $f=0.131463$, which is from the periodic region of the bifurcation diagram being non-extreme events region as well. No portion of the histogram is observed to surpass the extreme events qualifying \textit{red-dashed} threshold line towards the right. For the detailed numerical calculations, the other parameter values are considered as $a=-1.121$, $b=-0.6047$, $\beta=0.05$, and $\omega=0.265$.
	}  
	\label{fig3}
\end{figure*}
diagrams, the \textit{red-dashed} vertical line represents the extreme events qualifying threshold $H_{th}=m+6\mu$, where $m$ is the mean and $\mu$ is the standard deviation of the concerning data set. Figure \ref{fig3}(a) represents the histogram of the events $x_{max}$ corresponding to $f=0.131228$ in semi-log scale. This value of $f$ being from the non-extreme chaotic region of the bifurcation diagram, presented in Fig. \ref{fig1}(a), no portion of the histogram is observed to surpass the extreme events qualifying \textit{red-dashed} vertical threshold line towards the right. The histogram of the events $x_{max}$ in semi-log scale corresponding to $f=0.1314$ is portrayed in Fig. \ref{fig3}(b). This value of $f$ is from the EEs emerging chaotic region of the bifurcation diagram, shown in Fig. \ref{fig1}(a). A fragment of the histogram is noticed to exceed the EEs qualifying \textit{red-dashed} vertical threshold line toward the right. This portion \textit{au fond} corroborates the EEs. Figure \ref{fig3}(c) discerns the histogram of the events $x_{max}$ in semi-log scale regarding $f=0.131417$, which is also from the EEs emerging chaotic region of the bifurcation diagram, shown in Fig. \ref{fig1}(a). A portion of the histogram is observed to cross the EEs qualifying \textit{red-dashed} vertical line. This segment of the histogram endorses the EEs portion. The histogram of the events $x_{max}$ corresponding to $f=0.131463$ is displayed in semi-log scale in Fig. \ref{fig3}(d). This value of $f$ being from the non-extreme periodic region of the bifurcation diagram, presented in Fig. \ref{fig1}(a), no segment of the histogram is discerned to surpass the EEs qualifying \textit{red-dashed} vertical threshold line. All the numerical simulations are performed considering $10^7$ iterations and abjuring a transient of $10^5$.\\

\textbf{\textit{Probability distribution of the threshold excess values}:} A comprehensive exposition of the probability distributions of the threshold \textit{excess values} corresponding to $f=0.1314$ and $f=0.131417$,  
\begin{figure*}[!htbp]
		\centerline{
		\includegraphics[scale = 0.53]{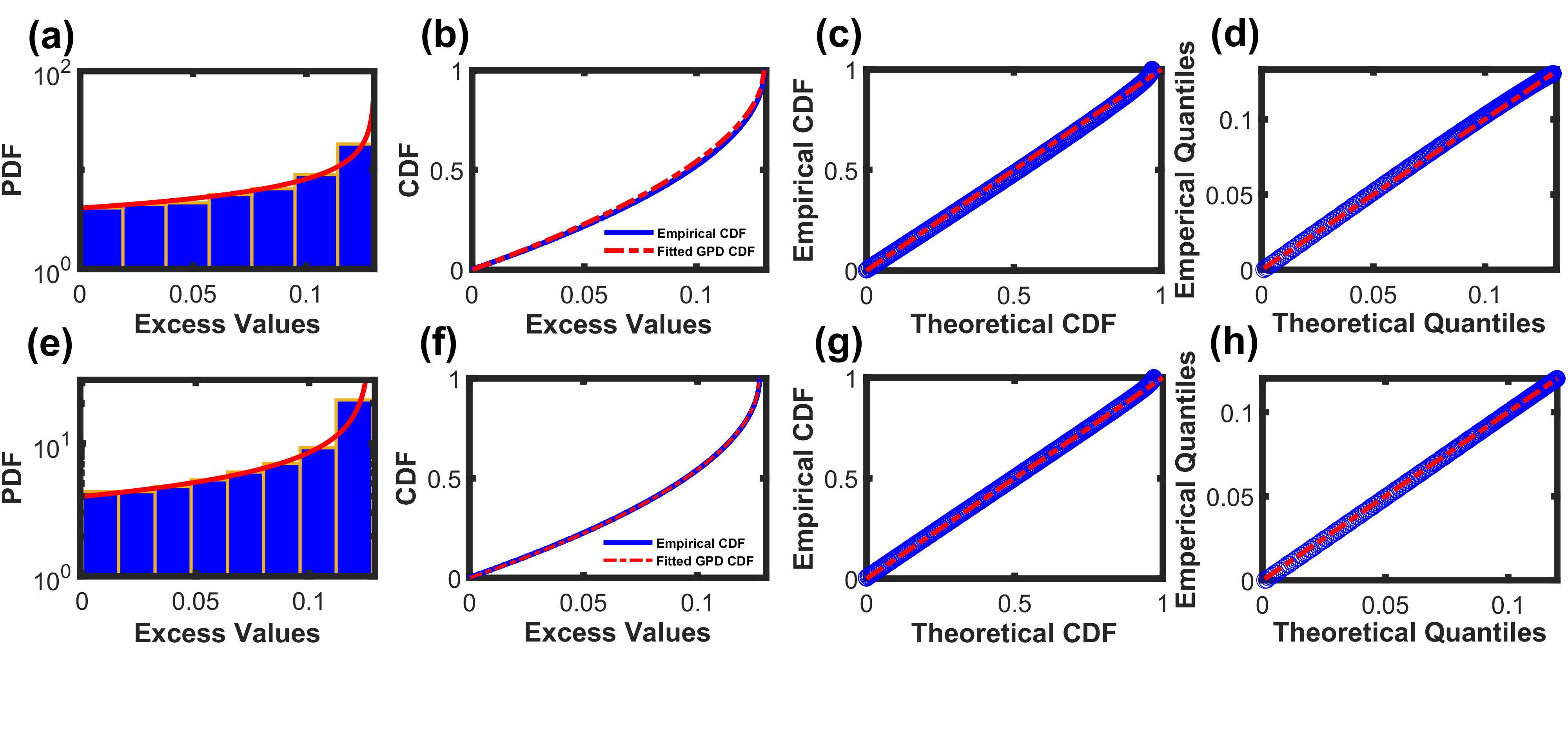}
		
				}
	\caption{\textbf{\textit{Statistical analysis of the sets of threshold excess values for different values of the bifurcation parameter $f$ of the system \eqref{eq1} concerning the events $x_{_{max}}$}:} (a) The fitted GPD distribution curve of the set of threshold excess values corresponding to $f=0.1314$ is depicted in \textit{red} color. The figure is presented in semi-log scale. The K-S statistic plot, P-P plot, and Q-Q plot of the set of threshold excess values for $f=0.1314$ regarding the GPD fitting are presented in (b)-(d), respectively. (e) The fitted GPD distribution curve of the set of threshold excess values for $f=0.131417$ is presented in \textit{red} color. The figure is delineated in semi-log scale. The K-S statistic plot, P-P plot, and Q-Q plot of the set of threshold excess values corresponding to $f=0.131417$ concerning the GPD fitting are displayed in (f)-(h). For the detailed numerical calculations, the other parameter values are considered as $a=-1.121$, $b=-0.6047$, $\beta=0.05$, and $\omega=0.265$. 
	}  
	\label{fig4}
\end{figure*} 
explicating the goodness of fit, is presented in this section and elucidated in Fig. \ref{fig4}. In the first panel, Fig. \ref{fig4}(a) displays the PDF (probability distribution function) of the \textit{threshold excess} values for $f=0.1314$, and Fig. \ref{fig4}(e) represents the PDF of the \textit{threshold excess} values corresponding to $f=0.131417$. Both the PDFs fit well with the GPD distribution, and their respective parameter values and the corresponding $95\%$ confidence intervals and the standard errors are represented in Table \ref {table1}. To 
\begin{table*} 	
	\centering
	\small
	\setlength{\tabcolsep}{0.05pt}
	\begin{tabular}{|c|c|c|c|c|c|}
		\hline
		\multicolumn{6}{|c|}{Statistics of the GPD distribution} \\\hline\hline
		Data & Bifurcation Parameter & Parameter & Estimate & Confidence Interval ($95\%$)& Standard Error \\ 
		\hline\hline
		& & $\alpha$ & $-1.86402$ & $[-2.00298, -1.79151]$ & $1.17470$ \\ 
		\cline{3-6}
		& $f = 0.1314$ &  &  &  &  \\ 
		\cline{3-6}
		& & $\hat{\gamma}$ & $0.24274$ & $[0.23330, 0.26084]$ & $0.15298$ \\
		\cline{2-6}
		Excess Value & & & & &\\
		\cline{2-6}
		& & $\alpha$ & $-1.97232$ & $[-2.01275, -1.88538]$ & $0.56160$ \\ 
		\cline{3-6}
		& $f = 0.131417$ &  &  &  &  \\ 
		\cline{3-6}
		& & $\hat{\gamma}$ & $0.25065$ & $[0.23960, 0.25579]$ & $0.07137$ \\ 
		\hline
	\end{tabular}
	\caption{Details of the GPD probability distribution curves concerning $f = 0.1314$ and $f = 0.131417$, respectively.}
	\label{table1}
\end{table*}
check how well this fit is, we did the Kolmogorov-Smirnov (K-S) test. In both cases the data sets (the sets of all \textit{threshold excess} values) concerning two different values of $f$ fail to relinquish the \textit{null hypothesis} (the respective data sets follow the GPD distribution) as far as the $95\%$ confidence interval is considered. This corroborates the good fitting of the PDFs with the GPD. The results of the K-S test are presented in Table \ref{table2}.
\begin{table*} 	
	\centering
	\small
	\setlength{\tabcolsep}{0.05pt}
	\begin{tabular}{|c|c|c|c|}
		\hline
		\multicolumn{4}{|c|}{K-S test result of GPD distributions of Excess Values } \\\hline\hline
		Data & Bifurcation Parameter & Parameter & Estimate \\ 
		\hline\hline
		& & $p$-value & $0.98375$ \\ 
		\cline{3-4}
		& $f = 0.1314$ & & \\ 
		\cline{3-4}
		& & $K$-$S$ Statistics & $0.02537$ \\
		\cline{2-4}
		Excess Value & & & \\
		\cline{2-4}
		& & $p$-value & $1.0$ \\ 
		\cline{3-4}
		& $f = 0.131417$ & &\\ 
		\cline{3-4}
		& & $K$-$S$ Statistics & $0.0056$ \\ 
		\hline
	\end{tabular}
	\caption{ K-S test results of the GPD distributions for the sets of Excess Values concerning $f = 0.1314$ and $f = 0.131417$, respectively. }
	\label{table2}
\end{table*}
The cumulative distribution function (CDF) calculated from the data set of the threshold \textit{excess values} being numerically simulated and the CDF obtained from the theoretically fitted GPD curve are presented in the \textit{blue} and \textit{red} curves, respectively, in the second panel; Fig. \ref{fig4}(b) represents for $f=0.1314$ and Fig. \ref{fig4}(f) represents for $f=0.131417$. The close alignment of the theoretically calculated CDF and the CDF calculated from the numerically generated data set affirms the good fit. For the visual comprehension of the goodness of fit, the probability-probability plot (P-P plot) is depicted in the third panel; Fig. \ref{fig4}(c) represents for $f=0.1314$, and Fig. \ref{fig4}(g) represents for $f=0.131417$. The quantile-quantile plot (Q-Q plot) is delineated in the fourth column, which also corroborates the visual confirmation of the good fitting; Fig. \ref{fig4}(d) represents the figure for $f=0.1314$, and Fig. \ref{fig4}(h) represents the figure for $f=0.131417$. 
The parameter values of the PDFs and their standard errors and the confidence intervals are calculated using the Jackknife method. The numerical simulations are performed considering $5 \times 10^5$ iterations and abjuring a transient of $10^5$.\\ 

\begin{figure*}[!htbp]
		\centerline{
		\includegraphics[scale = 0.53]{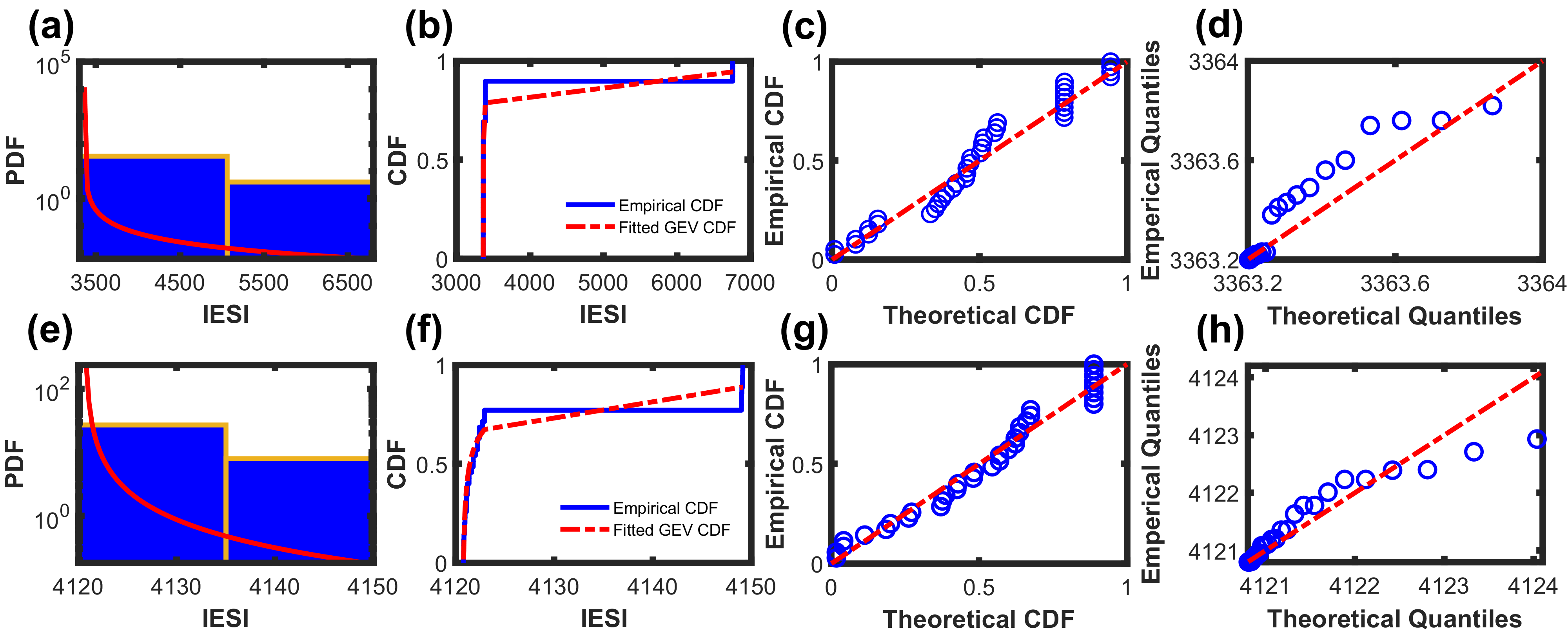}
		
				}
	\caption{\textbf{\textit{Statistical analysis of the sets of IESIs for different values of the bifurcation parameter $f$ of the system \eqref{eq1} concerning the events $x_{_{max}}$}:} (a) The fitted GEV distribution curve of the set of IESIs corresponding to $f=0.1314$ is depicted in \textit{red} color. The figure is presented in semi-log scale. The K-S statistic plot, P-P plot, and Q-Q plot of the set of IESIs for $f=0.1314$ regarding the GEV distribution fitting are presented in (b)-(d), respectively. (e) The fitted GEV distribution curve of the set of IESIs regarding $f=0.131417$ is shown in \textit{red} color. The figure is presented in semi-log scale. The K-S statistic plot, P-P plot, and Q-Q plot of the set of IESIs corresponding to $f=0.131417$ concerning the GEV distribution fitting are portrayed in (f)-(h), respectively. For the detailed numerical calculations, the other parameter values are considered as $a=-1.121$, $b=-0.6047$, $\beta=0.05$, and $\omega=0.265$.
	}  
	\label{fig5}
\end{figure*}

\textbf{\textit{Probability distribution of the inter-extreme-spike-intervals}:} A perceptive illustration of the probability distributions of the IESIs concerning $f=0.1314$ and $f=0.131417$, interpreting the goodness of fit, is presented in this section and explicated in Fig. \ref{fig5}.
In the first panel, Fig. \ref{fig5}(a) represents the PDF of IESIs corresponding to $f=0.1314$, and Fig. \ref{fig5}(b) shows the PDF of the IESIs regarding $f=0.131417$. Both the PDFs fit well with the GEV distribution, and their respective parameter values, $95\%$ confidence intervals, and standard errors are presented in Table \ref{table3}. To check the goodness 
\begin{table*} 	
	\centering
	\small
	\setlength{\tabcolsep}{0.05pt}
	\begin{tabular}{|c|c|c|c|c|c|}
		\hline
		\multicolumn{6}{|c|}{Statistics of the GEV distribution} \\\hline\hline
		Data & Bifurcation Parameter & Parameter & Estimate & Confidence Interval ($95\%$)& Standard Error \\ 
		\hline\hline
		& & $\gamma$ & $3.32609$ & $[0.541451, 6.1107321]$ & $1.420735$ \\ 
		\cline{3-6}
		& $f = 0.1314$ & $\beta$ & $0.828916$ & $[-16.669497, 18.327330]$ & $8.927762$ \\ 
		\cline{3-6}
		& & $\alpha$ & $3363.45$ & $[3359.151476, 3367.744200]$ & $2.192021$ \\
		\cline{2-6}
		IESI & & & & &\\
		\cline{2-6}
		& & $\gamma$ & $2.152240$ & $[1.006208, 3.298271]$ & $0.584710$ \\ 
		\cline{3-6}
		& $f = 0.131417$ & $\beta$ & $0.624453$ & $[0.020658, 1.228249]$ & $0.308059$ \\ 
		\cline{3-6}
		& & $\alpha$ & $4121.076086$ & $[4120.710368, 4121.441804]$ & $0.186591$ \\ 
		\hline
	\end{tabular}
	\caption{Details of the GEV distributions of IESIs concerning the bifurcation parameter values $f = 0.1314$ and $f = 0.131417$.}
	\label{table3}
\end{table*}
of fit, we performed the K-S test. In both cases the sets of IESIs corresponding to $f=0.1314$ and $f=0.131417$ fail to relinquish the \textit{null hypothesis} (the respective data sets follow the GEV distribution) so far as the $95\%$ confidence interval is concerned. This confirms the good fitting of the PDFs with the GEV distribution. The results of the K-S tests presented in Table \ref{table4}. The CDF calculated from the numerically generated data set (IESIs)
\begin{table*} 	
	\centering
	\small
	\setlength{\tabcolsep}{0.05pt}
	\begin{tabular}{|c|c|c|c|}
		\hline
		\multicolumn{4}{|c|}{K-S test result of GEV distributions of IESIs } \\\hline\hline
		Data & Bifurcation Parameter & Parameter & Estimate \\ 
		\hline\hline
		& & $p$-value & $0.4715$ \\ 
		\cline{3-4}
		& $f = 0.1314$ & & \\ 
		\cline{3-4}
		& & $K$-$S$ Statistics & $0.1314$ \\
		\cline{2-4}
		IESI & & & \\
		\cline{2-4}
		& & $p$-value & $0.6891$ \\ 
		\cline{3-4}
		& $f = 0.131417$ & &\\ 
		\cline{3-4}
		& & $K$-$S$ Statistics & $0.1162$ \\ 
		\hline
	\end{tabular}
	\caption{ K-S test results of the GEV distributions for the sets of IESIs concerning $f = 0.1314$ and $f = 0.131417$, respectively. }
	\label{table4}
\end{table*}
and the CDF calculated from the theoretically fitted GEV distribution function are presented in \textit{blue} and \textit{red} curves, respectively, in the second panel; Fig. \ref{fig5}(b) corresponds to $f=0.1314$, and Fig. \ref{fig5}(f) corresponds to $f=0.131417$. The close alignment of the numerically generated CDF and the theoretically generated CDF affirms the quality of the good fit. For the visual discernment of the good fitting, the P-P plot and the Q-Q plot are delineated in the third and fourth panels, respectively. Figure \ref{fig5}(c) represents the P-P plot corresponding to $f=0.1314$, and Fig. \ref{fig5}(g) depicts the P-P plot concerning $f=0.131417$. Figure \ref{fig5}(d) represents the Q-Q plot regarding $f=0.1314$, and Fig. \ref{fig5}(h) depicts the Q-Q plot corresponding to $f=0.131417$. The parameter values of the PDFs and their standard errors and the confidence intervals are calculated using the Jackknife method. The numerical simulations are performed considering $1.5 \times 10^7$ iterations and abjuring a transient of $10^5$.\\

\subsection{Statistical analysis concerning the events $x_{min}$}
\label{sec:leve5ii}
This section represents a careful scrutinization regarding various statistical perspectives of the events $x_{min}$ for different values of $f$ of the system \eqref{eq1}. At first a detailed description of the histograms for different values of $f$ is interpreted. Secondly, a comprehensive prescription of the probability distributions of the threshold \textit{excess values} (the absolute value of the difference between the EE and the EE qualifying threshold $H_{th}$) is portrayed, and lastly, how the \textit{inter-extreme-spike-intervals} (IESIs) behave statistically concerning the probability distribution for different values of $f$ is elucidated.\\

\textbf{\textit{Histogram presentation}:} A comprehensive and circumstantial description of the histograms of the events $x_{min}$ regarding different values of $f$ is  
\begin{figure*}[!htbp]
		\centerline{
		\includegraphics[scale = 0.56]{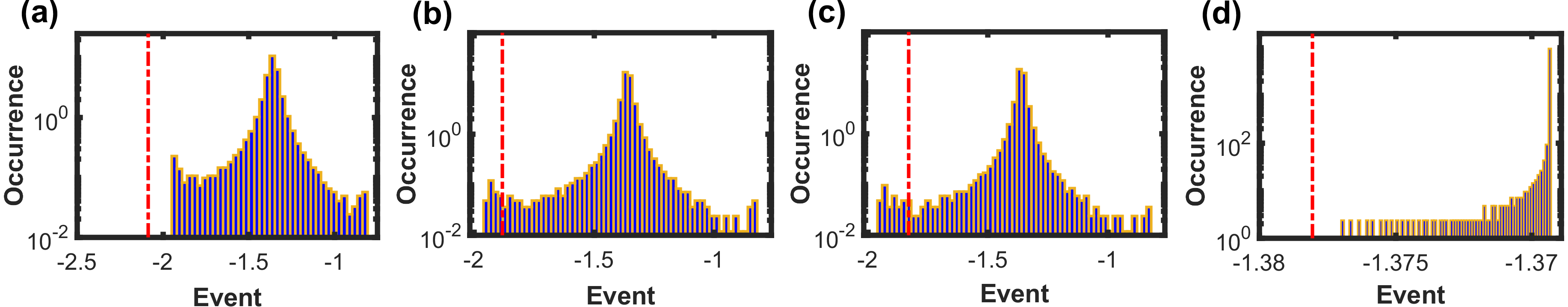}
		
				}
	\caption{\textbf{\textit{Histogram of the events $x_{_{min}}$ for different values of the bifurcation parameter $f$ of the system \eqref{eq1} in semi-log scale}:} The \textit{green-dashed} vertical line represents the extreme events qualifying threshold $H_{th}=m-6\mu$, where $m$ is the mean and $\mu$ is the standard deviation of the respective set of events for different values of $f$. (a) Histogram of the events corresponding to the bifurcation parameter $f=0.131231$, which is from the non-extreme chaotic region of the bifurcation diagram. No portion of the histogram is noticed to cross the extreme events qualifying \textit{green-dashed} threshold line towards the left. (b) Histogram of the events concerning $f=0.131393$ being from the extreme events qualifying region of the bifurcation diagram. A portion of the histogram is discerned to surpass the extreme events qualifying \textit{green-dashed} vertical threshold line towards the left, which indeed corroborates the extreme events portion. (c) Histogram of the events for $f=0.131414$, which is also from the extreme events qualifying region of the bifurcation diagram. A portion of the histogram is observed to exceed the extreme events qualifying \textit{green-dashed} vertical line towards the left, which is basically the extreme events portion. (d) Histogram of the events regarding $f=0.131471$ being from the non-extreme periodic region of the bifurcation diagram. No portion of the histogram is seen to cross the extreme events qualifying \textit{green-dashed} vertical line towards the left. For the detailed numerical calculations, the other parameter values are considered as $a=-1.121$, $b=-0.6047$, $\beta=0.05$, and $\omega=0.265$.
	}  
	\label{fig6}
\end{figure*}
delineated in this segment. The \textit{green-dashed} vertical line in each histogram figure illustrates the EEs qualifying threshold $H_{th}=m-6\mu$, where $m$ is the mean and $\mu$ is the standard deviation of the respective data set. Figure \ref{fig6}(a) depicts the histogram of the events $x_{min}$ corresponding to $f=0.131231$, which, being from the non-extreme chaotic region of the bifurcation diagram, presented in Fig. \ref{fig1}(a), no portion of the histogram is noticed to cross the \textit{green-dashed} EEs qualifying vertical threshold line towards the left. The histogram of the events $x_{min}$ concerning $f=0.131393$ is displayed in Fig. \ref{fig6}(b). This value of $f$ being from the EEs emerging chaotic region of the bifurcation diagram, delineated in Fig. \ref{fig1}(a), a portion of the histogram is discerned to exceed the EEs qualifying \textit{green-dashed} vertical line towards the left. This portion of the histogram basically represents the EEs. Figure \ref{fig6}(c) elucidates the histogram of the events $x_{min}$ regarding $f=0.131414$, which is also from the EEs emerging chaotic region of the bifurcation diagram, displayed in Fig. \ref{fig1}(a), a segment of the histogram is noticed to surpass the EEs qualifying \textit{green-dashed} vertical line towards the left. This portion of the histogram indeed depicts the EEs. The histogram of the events $x_{min}$ corresponding to $f=0.131471$ is delineated in Fig. \ref{fig6}(c). This value of $f$ being from the non-extreme periodic region of the bifurcation diagram, presented in Fig. \ref{fig1}(a), no portion of the histogram is noticed to transcend the EEs qualifying \textit{green-dashed} vertical threshold line towards the left. All the numerical simulations are performed considering $10^7$ iterations and abjuring a transient of $10^5$.\\

\begin{figure*}[!htbp]
		\includegraphics[scale = 0.53]{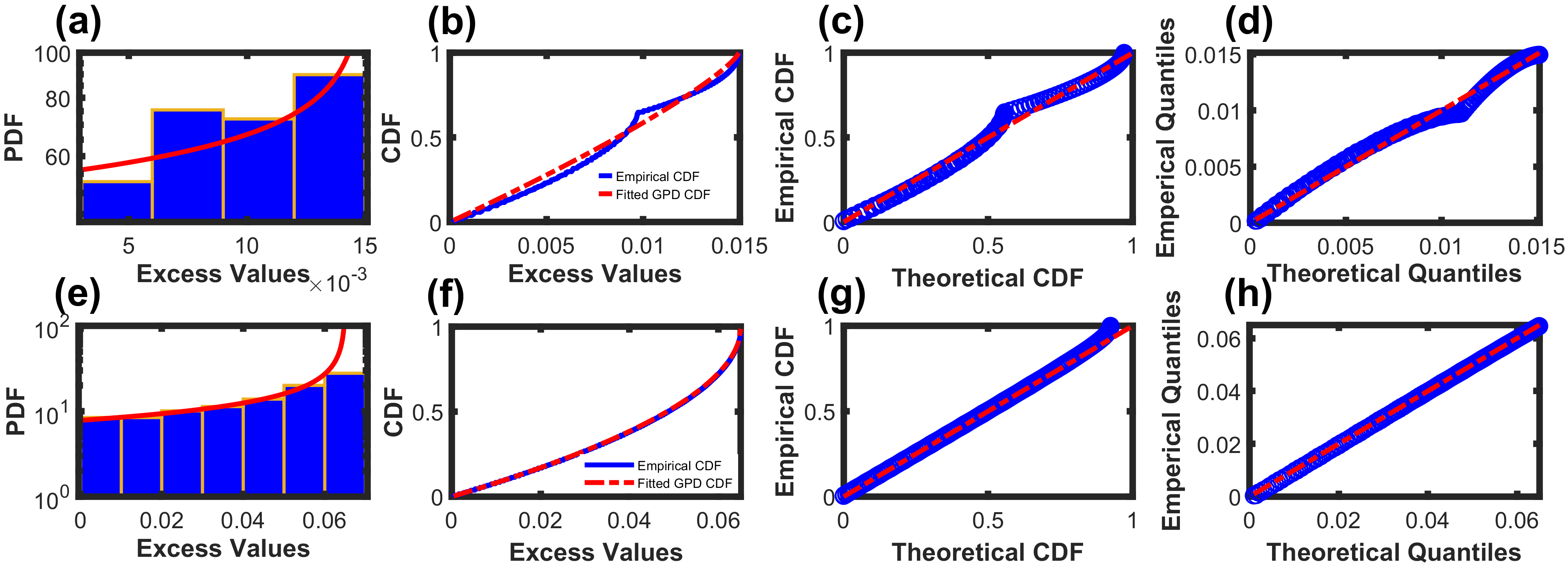}
		
	\caption{\textbf{\textit{Statistical analysis of the sets of threshold excess values for different values of the bifurcation parameter $f$ of the system \eqref{eq1} concerning the events $x_{_{min}}$}:} (a) The fitted GPD curve of the set of the threshold excess values for $f=0.131393$ is depicted in \textit{red} color. The figure is portrayed in semi-log scale. The K-S statistic, P-P plot, and Q-Q plot of the set of the threshold excess values for $f=0.131393$ regarding the GPD fitting are delineated in (b)-(d), respectively. (e) The fitted GPD curve corresponding to $f=0.131414$ is shown in \textit{red} color. The figure is presented in semi-log scale. The K-S statistic, P-P plot, and Q-Q plot of the set of threshold excess values for $f=0.131414$ concerning the GPD fitting are displayed in (f)-(h), respectively. For the detailed numerical calculations, the other parameter values are considered as $a=-1.121$, $b=-0.6047$, $\beta=0.05$, and $\omega=0.265$.
	}  
	\label{fig8}
\end{figure*}

\textbf{\textit{Probability distribution of the threshold excess values}:} A cognitive illustration of the probability distributions of the \textit{threshold excess} values concerning $f=0.131393$ and $f=0.131414$, scrutinizing the goodness of fit, is prescribed in this section and interpreted in Fig. \ref{fig8}. In the first panel, Fig. \ref{fig8}(a) depicts the PDF of the \textit{threshold excess} values corresponding to $f=0.131393$, and Fig. \ref{fig8}(e) shows the PDF of the \textit{threshold excess} 
values corresponding to $f=0.131414$. Both the PDFs fit well with the GPD and their concerning parameter values and the respective $95\%$ confidence intervals and the standard errors are enlisted in Table \ref{table5}. To test the goodness of fit, K-S test is performed. Both the data sets
\begin{table*} 	
	\centering
	\small
	\setlength{\tabcolsep}{0.05pt}
	\begin{tabular}{|c|c|c|c|c|c|}
		\hline
		\multicolumn{6}{|c|}{Statistics of the GPD distribution} \\\hline\hline
		Data & Bifurcation Parameter & Parameter & Estimate & Confidence Interval ($95\%$)& Standard Error \\ 
		\hline\hline
		& & $\alpha$ & $-1.24770$ & $[-1.30719, -1.21225]$ & $0.33042$ \\ 
		\cline{3-6}
		& $f = 0.131393$ &  &  &  &  \\ 
		\cline{3-6}
		& & $\hat{\gamma}$ & $0.01866$ & $[0.01813, 0.01954]$ & $0.00494$ \\
		\cline{2-6}
		Excess Value & & & & &\\
		\cline{2-6}
		& & $\alpha$ & $-1.96263$ & $[-1.99920, -1.49314]$ & $1.86919$ \\ 
		\cline{3-6}
		& $f = 0.131414$ &  &  &  &  \\ 
		\cline{3-6}
		& & $\hat{\gamma}$ & $0.12776$ & $[0.09720, 0.13014]$ & $0.12168$ \\ 
		\hline
	\end{tabular}
	\caption{Details of the GPD probability distribution curves concerning $f = 0.131393$ and $f = 0.131414$, respectively.}
	\label{table5}
\end{table*}
(the sets of \textit{threshold excess} values) regarding two different values of $f$ fail to relinquish the \textit{null hypothesis} (the respective data set follows the GPD) so far as the $95\%$ confidence interval is considered. This confirms the good fitting of the PDFs with the GPD. The
\begin{table*} 	
	\centering
	\small
	\setlength{\tabcolsep}{0.05pt}
	\begin{tabular}{|c|c|c|c|}
		\hline
		\multicolumn{4}{|c|}{K-S test result of GPD distributions of Excess Values } \\\hline\hline
		Data & Bifurcation Parameter & Parameter & Estimate \\ 
		\hline\hline
		& & $p$-value & $0.13645$ \\ 
		\cline{3-4}
		& $f = 0.131393$ & & \\ 
		\cline{3-4}
		& & $K$-$S$ Statistics & $0.07952$ \\
		\cline{2-4}
		Excess Value & & & \\
		\cline{2-4}
		& & $p$-value & $1.0$ \\ 
		\cline{3-4}
		& $f = 0.131417$ & &\\ 
		\cline{3-4}
		& & $K$-$S$ Statistics & $0.00946$ \\ 
		\hline
	\end{tabular}
	\caption{ K-S test results of the GPD distributions for the sets of Excess Values concerning $f = 0.131393$ and $f = 0.131414$, respectively. }
	\label{table6}
\end{table*}
results of the K-S test are enlisted in Table \ref{table6}. In the second panel the CDF calculated from the numerically generated data set (the set of \textit{threshold excess} values) is presented in the \textit{blue} curve, and the CDF being calculated theoretically from the fitted GPD curve is depicted in the \textit{red} curve; specifically, Fig. \ref{fig8}(b) is the case study for $f=0.131393$, and Fig. \ref{fig8}(f) is the case study for $f=0.131414$. The close alignment of the numerically calculated CDF and the theoretically calculated CDF confirms the good fit. For the visual recognition of the well fit, the P-P plots are presented in the third panel; Fig. \ref{fig8}(c) is the P-P plot for $f=0.131393$, and Fig. \ref{fig8}(g) is the P-P plot for $f=0.131414$. The Q-Q plot is also another tool to affirm visually the good fit, being presented in the fourth panel; Fig. \ref{fig8}(d) is the Q-Q plot for $f=0.131393$, and Fig. \ref{fig8}(h) is the Q-Q plot for $f=0.131414$. The parameter values of the PDFs and their standard errors and the confidence intervals are calculated using the Jackknife method. The numerical simulations are performed considering $5 \times 10^5$ iterations and abjuring a transient of $10^5$.\\

\begin{figure*}[!htbp]
		\centerline{
		\includegraphics[scale = 0.535]{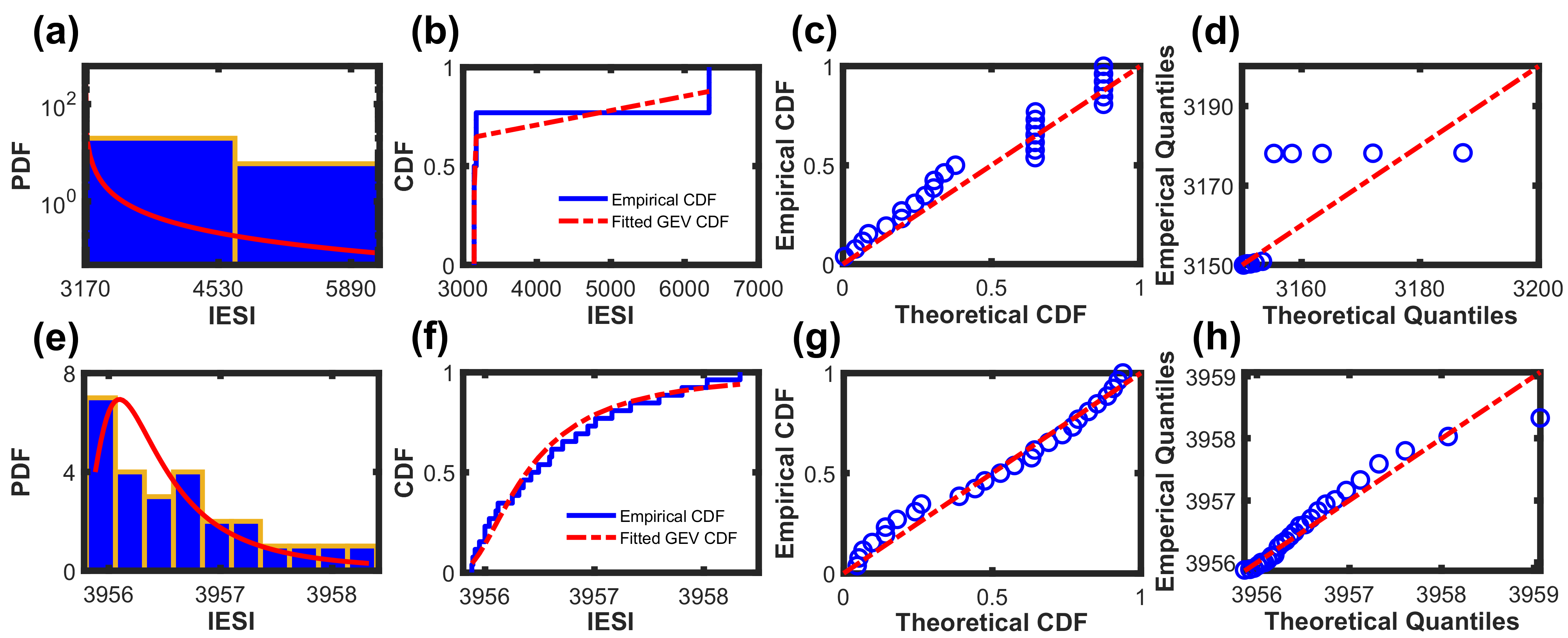}
		
				}
	\caption{\textbf{\textit{Statistical analysis of the sets of IESIs for different values of the bifurcation parameter $f$ of the system \eqref{eq1} concerning the events $x_{_{min}}$}:} (a) The fitted GEV distribution curve of the set of IESIs for $f=0.131393$ is displayed. The figure is presented in semi-log scale. The K-S statistic, P-P plot, and Q-Q plot of the set of IESIs corresponding to $f=0.131393$ concerning the GEV distribution fitting are portrayed in (b)-(d), respectively. (e) The fitted GEV distribution curve of the set of IESIs corresponding to $f=0.131414$ is shown in \textit{red} color. The figure is presented in semi-log scale. The K-S statistic plot, P-P plot, and Q-Q plot of the set of IESIs for $f=0.131414$ concerning the GEV distribution fitting are displayed in (f)-(h), respectively. For the detailed numerical calculations, the other parameter values are considered as $a=-1.121$, $b=-0.6047$, $\beta=0.05$, and $\omega=0.265$.
	}  
	\label{fig9}
\end{figure*}

\textbf{\textit{Probability distribution of the inter-extreme-spike-intervals}:} A perspicacious description of the probability distributions of the IESIs regarding the two different parameter values, $f=0.131393$ and $f=0.131414$, discerning how well the fit is, is presented in this section and elucidated in Fig. \ref{fig9}. In the first panel, Fig. \ref{fig9}(a) displays the PDF of IESIs corresponding to $f=0.131393$, and Fig. \ref{fig9}(e) depicts the PDF of the 
IESIs corresponding to $f=0.131414$. Both the PDFs fit well with the GEV distribution, and their concerning parameter values, $95\%$ confidence intervals, and standard errors are presented in Table \ref{table7}. For the scrutinization of 
\begin{table*} 	
	\centering
	\small
	\setlength{\tabcolsep}{0.05pt}
	\begin{tabular}{|c|c|c|c|c|c|}
		\hline
		\multicolumn{6}{|c|}{Statistics of the GEV distribution} \\\hline\hline
		Data & Bifurcation Parameter & Parameter & Estimate & Confidence Interval ($95\%$)& Standard Error \\ 
		\hline\hline
		& & $\gamma$ & $3.934460$ & $[-0.850424, 8.719344]$ & $2.441267$ \\ 
		\cline{3-6}
		& $f = 0.131393$ & $\beta$ & $4.232853$ & $[-234.534592, 243.000298]$ & $121.820125$ \\ 
		\cline{3-6}
		& & $\alpha$ & $3150.804089$ & $[3102.730408, 3198.877769]$ & $24.527388$ \\
		\cline{2-6}
		IESI & & & & &\\
		\cline{2-6}
		& & $\gamma$ & $0.436197$ & $[-0.293714, 1.166109]$ & $0.372404$ \\ 
		\cline{3-6}
		& $f = 0.131414$ & $\beta$ & $0.390071$ & $[0.159962, 0.620181]$ & $0.117403$ \\ 
		\cline{3-6}
		& & $\alpha$ & $3956.227277$ & $[3955.924799, 3956.529754]$ & $0.154325$ \\ 
		\hline
	\end{tabular}
	\caption{Details of the GEV distributions of IESIs concerning the bifurcation parameter values $f = 0.131393$ and $f = 0.131414$.}
	\label{table7}
\end{table*}
the goodness of fit, the K-S test is performed. In both the cases concerning $f=0.131393$ and $f=0.131414$, the sets of IESIs fail to relinquish the \textit{null hypothesis} (the respective data sets follow the GEV distribution) as far as the $95\%$ confidence interval is considered. This, indeed, corroborates the good fittings of the PDFs with the GEV distribution. The results of the K-S tests are presented in Table \ref{table8}. In the second panel, the CDF calculated 
\begin{table*} 	
	\centering
	\small
	\setlength{\tabcolsep}{0.05pt}
	\begin{tabular}{|c|c|c|c|}
		\hline
		\multicolumn{4}{|c|}{K-S test result of GEV distributions of IESIs } \\\hline\hline
		Data & Bifurcation Parameter & Parameter & Estimate \\ 
		\hline\hline
		& & $p$-value & $0.5780$ \\ 
		\cline{3-4}
		& $f = 0.131393$ & & \\ 
		\cline{3-4}
		& & $K$-$S$ Statistics & $0.1469$ \\
		\cline{2-4}
		IESI & & & \\
		\cline{2-4}
		& & $p$-value & $0.9592$ \\ 
		\cline{3-4}
		& $f = 0.131414$ & &\\ 
		\cline{3-4}
		& & $K$-$S$ Statistics & $0.0940$ \\ 
		\hline
	\end{tabular}
	\caption{ K-S test results of the GEV distributions for the sets of IESIs concerning $f = 0.131393$ and $f = 0.131414$, respectively. }
	\label{table8}
\end{table*}
from the numerically generated data set (the set of IESIs) and the CDF calculated from the theoretically fitted GEV distribution curve are presented in \textit{blue} and \textit{red} colors, respectively; Fig. \ref{fig9}(b) represents the case study corresponding to $f=0.131393$, and Fig. \ref{fig9}(f) represents the case study corresponding to $f=0.131414$. The quality of good fittings is being appraised by the close alignments of the numerically generated CDF and the theoretically generated CDF. For the visual affirmation of the good fittings, the P-P plot and Q-Q plot are presented in the third and in the fourth panels, respectively; Fig. \ref{fig9}(c) represents the P-P plot corresponding to $f=0.131393$, Fig. \ref{fig9}(g) depicts the P-P plot corresponding to $f=0.131414$, Fig. \ref{fig9}(d) displays the Q-Q plot regarding $f=0.131393$, and Fig. \ref{fig9}(h) represents the Q-Q plot for $f=0.131414$. The parameter values of the PDFs and their standard errors and the confidence intervals are calculated using the Jackknife method. The numerical simulations are performed considering $1.1 \times 10^7$ iterations and abjuring a transient of $10^5$.\\
\section{\label{sec:level5}Conclusions}

This study holistically spellbound us to investigate the dynamics of the MLC circuit as the amplitude of the externally applied periodic forcing is varied within a certain range. Interestingly, we found that the EEs emerge in the system concerning two different \textit{events} being defined. We also exposed the dynamical synergy responsible for the genesis of EEs in the system from three different perspectives: the prevalence of a \textit{force field} due to the presence of an externally applied periodic force, the stable and unstable manifolds decomposition concerning slow-fast dynamics, and Floquet multipliers. The considered MLC circuit variant in this study is a nonautonomous system. To compute the stable and unstable manifolds, the nonautonomous system is transformed to an autonomous one, preserving the system dynamics, and the dynamical mechanisms behind the observed phenomena are explained. This direction of explaining the nonlinear dynamical phenomena of different nonautonomous dynamical systems under the effect of external forcing being applied might be an engrossing and potential future research avenue. The explanation of the dynamical synergy being exposed because of the perception of the \textit{force field} being prevalent due to the presence of the external forcing in the system might be another engrossing research direction. The mathematical calculations of the magnitude and the direction of the \textit{force field} on the concerning surface that create the significant low and high intensity \textit{force field} responsible for the large deflection of the chaotic trajectory remain a challenging future research avenue. The stable and unstable manifolds calculation regarding Floquet multipliers specifically for nonautonomous systems under the application of the externally applied periodic force and how the manifolds affect the relevant phase space for the large excursion of the chaotic trajectory residing on the space claims more meticulous exploration. We also analyzed EEs statistically to confirm the rare occurrences using EVT. The threshold excess values follow the GPD, and the IESIs follow the GEV distribution. It remains a challenging and potential future research area to unveil the characteristics of the non-Gaussian probability distributions concerning the local stable and unstable manifolds, showing how the nature of the distributions changes when the chaotic trajectory deflects largely, showing the signature of EEs while wandering within a comparatively small region of the phase space. In conclusion, we hope that the findings of our study will be benevolent in any future research concerning the pertinent directions.

\section{\label{sec:level6}Appendix}

The system \eqref{eq1} is nonautonomous; initially we transform it to an autonomous one by introducing the variable $z=\omega t$.
Therefore, the system reduces to
\begin{equation}
	\begin{aligned}
		\dot{x} &= f \sin z - x - y - g(x), \\
		\dot{y} &= \beta x, \\
		\dot{z} &= \omega
	\end{aligned}
	\label{eq2}
\end{equation}
where
\[
g(x) =
\begin{cases}
	bx + (a - b), & x > 1, \\
	ax,           & |x| \leq 1, \\
	bx - (a - b), & x < -1
\end{cases}
\]

The newly introduced variable $z=\omega t$ behaves like a phase variable. The period of the externally applied periodic forcing is $T = \frac{2\pi}{\omega}$.\\

\subsection{Calculation of the Floquet multipliers and the construction of the stable and unstable manifolds}
\label{sec:leve16i}

The Jacobian of \eqref{eq2} is

\[
J =
\begin{pmatrix}
	\frac{\partial \dot{x}}{\partial x} & \frac{\partial \dot{x}}{\partial y} & \frac{\partial \dot{x}}{\partial z} \\
	\frac{\partial \dot{y}}{\partial x} & \frac{\partial \dot{y}}{\partial y} & \frac{\partial \dot{y}}{\partial z} \\
	\frac{\partial \dot{z}}{\partial x} & \frac{\partial \dot{z}}{\partial y} & \frac{\partial \dot{z}}{\partial z}
\end{pmatrix}
\]

\textit{i.e.,}

\[
J =
\begin{pmatrix}
	-1 - g'(x) & -1 & f \cos z \\
	\beta      & 0  & 0 \\
	0          & 0  & 0
\end{pmatrix}
\]

where

\[
g'(x)=
\begin{cases}
	b, & x>1,\\
	a, & |x| \leq 1,\\
	b, & x<-1,
\end{cases}
\]

Suppose the fundamental matrix is $\Phi(t)$. Then 

\begin{equation}
	\dot{\Phi}(t)=J\Phi(t), \quad \Phi(0)=I
\end{equation}

The monodromy matrix is calculated as $M=\Phi(T)$, where $T$ is the period of the external forcing. The eigenvalues of the monodromy matrix $\lambda_{i}, i=1,2,3$, are the Floquet multipliers. If the absolute value of the Floquet multiplier is less than $1$, then it is stable, and the corresponding eigenvector infers the stable direction. Contrapositively, if the absolute value of the Floquet multiplier is greater than $1$, it represents an unstable one, and the respective eigenvector infers the unstable direction.\\

Suppose $X_{p}(t)=(x_{p}(t),y_{p}(t),z_{p}(t))$ is a periodic solution (\textit{i.e.,} a periodic orbit) of period $T$.\\

\textit{i.e.,} $X_{p}(t+T)=X_{p}(t)$\\

Let $X_{p}(0)$ be any point on the periodic orbit. \\

\textbf{\textit{Attracting (or stable) manifold construction}:} The initial condition is considered as $X_{0}=X_{p}(0)+\epsilon v_{s}$, where $v_{s}$ denotes the eigenvector corresponding to the stable Floquet multiplier, and $0 < \epsilon \ll 1$. Now, integrating the system \eqref{eq2} \textit{forward in time}, the solution $X=(x,y,z)$ is obtained. This bunch of trajectories forms the stable manifold.\\
\textbf{\textit{Repelling (or unstable) manifold construction}:} The initial condition is considered as $X_{0}=X_{p}(0)+\epsilon v_{u}$, where $v_{u}$ denotes the eigenvector corresponding to the unstable Floquet multiplier, and $0 < \epsilon \ll 1$. Now, integrating the system \eqref{eq2} \textit{backward in time}, the solution $X=(x,y,z)$ is obtained. This bunch of trajectories forms the unstable manifold.\\

\subsection{Calculation of the stable and unstable manifolds concerning slow-fast dynamics}
\label{sec:leve16ii}

The values of the parameters $\beta$ and $\omega$ are considered as $0.05$ and $0.265$; both are less than $1$. But $\beta \ll 1$. So, the system \eqref{eq2} can be contemplated as a slow-fast dynamical system, where $x$ is the fast variable, $y$ is the slow variable, and the variable $z$ behaves as slow constant drift.\\

The critical manifold:\\
\[
S = \left\{ (x, y, z) : y = f \sin z - x - g(x) \right\}\\
\]\\

From the system \eqref{eq2}, we obtain\\
\[
\frac{\partial \dot{x}}{\partial x} = -1 - g'(x)\\
\]\\

The stable (or attracting) segment of the critical manifold is defined by $\frac{\partial \dot{x}}{\partial x}<0$, and the unstable (or repelling) segment of the critical manifold is defined by $\frac{\partial \dot{x}}{\partial x}>0$.\\

The folds occur at the switching points $x=\pm 1$, and the fold curves are calculated by substituting the switching points at $y=f\sin z - x - g(x)$ (slow manifold).



%
\bibliographystyle{unsrt}       


\end{document}